\documentclass{aa}

\usepackage{graphicx}
\usepackage{txfonts}
\usepackage{lipsum}
\usepackage{subcaption}
\usepackage{lscape}
\usepackage{placeins}
\usepackage{hyperref}
\usepackage{natbib}
\usepackage{csquotes}
\usepackage{xcolor}
\usepackage{cuted}

\graphicspath{{./figures/}}
\bibpunct{(}{)}{;}{a}{}{,}
\hypersetup{
    colorlinks=true,
    linkcolor=blue,
    citecolor=blue,
    filecolor=magenta,      
    urlcolor=blue,
}

\newcommand{\editthree}[1]{#1}

\begin{document}

\title{Deep radiative zones affect giant planet cooling and internal structure: Implications for exoplanet characterisation}
\titlerunning{Deep radiative zones in giant planets}

\author{
    Simon Müller\inst{1}
    \and
    Ravit Helled\inst{1}
}
\authorrunning{Müller \& Helled}

\institute{
    Department of Astrophysics, University of Zürich, \\
    Winterthurerstrasse 190, 8057 Zürich, Switzerland \\
    \email{simonandres.mueller@uzh.ch}
}

\date{Received Month Day, Year}

\abstract
% context heading
{The radiative opacity plays a critical role in shaping both the thermal evolution and interior structure of giant planets. At temperatures of $\sim$2,000 K, there is a hydrogen-transparency region that creates a window of reduced opacity, which can give rise to detached, deep radiative zones nestled between two convective zones. This \editthree{local minimum in the opacity could be further deepened} by the depletion of alkali metals. While the existence of such zones has been explored in the context of Jupiter and Saturn, their influence on the evolution and characterisation of cold to warm giant exoplanets remains unstudied.}
% aims heading
{We investigate how opacity windows and the resulting deep radiative zones affect the cooling, radius evolution, and characterisation of the interiors and atmospheres of giant exoplanets.} 
% methods heading
{We computed thermal evolution models for {cold to} warm Jupiters spanning masses of 0.3 to {4.0 M$_{\rm{J}}$, envelope metallicities from one to ten times solar, equilibrium temperatures of 100 to 800 K}, and a parametrised reduction in the radiative opacity.}
% results heading
{Detached deep radiative zones develop in moderately irradiated Jupiters older than a few gigayears even with unmodified opacities, and earlier and more extensively when the opacity is reduced. The age and equilibrium temperature at which a detached deep radiative zone appears depends on the planetary mass, envelope metallicity, and opacity, with metal-enrichment suppressing deep radiative zones at low equilibrium temperatures but promoting them at higher ones. A deep opacity window accelerates the planetary cooling, reducing the predicted radii by up to 5\% and interior temperatures by several tens of percent. This translates to a $\sim$ 10 percentage point difference in the inferred bulk metallicity.}
% conclusions heading
{Deep radiative zones are likely common in warm giant exoplanets and could decouple atmospheric composition from bulk interior composition, complicating the interpretation of atmospheric observations. We suggest that the opacity treatment introduces significant uncertainties in atmospheric and interior characterisation.}


\maketitle
\nolinenumbers  % remove line numbers for arXiv

\section{Introduction}\label{sec:introduction}
Giant planets play a key role in our understanding of planetary systems. Efforts to detect and characterise giant exoplanets are ongoing. However, the characterisation of giant exoplanets strongly depends on their internal structure and thermal evolution. For decades, the standard model of gas giant interiors, exemplified by Jupiter, assumed a largely isentropic, fully convective state. Recent gravity field measurements from the Juno mission \citep{Bolton2017,Folkner2017,2020GeoRL..4786572D} have challenged this paradigm \citep{Wahl2017,Debras2019,2022A&A...662A..18M,2022PSJ.....3..185M,howard_jupiters_2023}. In addition, it became clear that there is tension between the atmospheric metallicity measured by the Galileo probe \citep{2000JGR...10515061M} and the predicted envelope metallicity from interior models \citep{2023A&A...680L...2H,cozza_denser_2026}. \citet{muller_can_2024} suggested that this tension can be resolved if Jupiter has a {detached deep radiative zone in the outer envelope located between two convection zones caused by a drop in opacity for local temperatures of $\sim$ 2,000 K}. This opacity window, primarily driven by the depletion of alkali metals and the behaviour of molecular hydrogen \citep{1994Icar..112..354G,1994Icar..112..337G,Guillot2004,Freedman2008,siebenaler_conditions_2025}, suggests that Jupiter has a deep radiative zone at a pressure of $\sim 10^3$ bar and that the metallicity of its atmosphere does not represent the metallicity of the deeper interior. {Furthermore, as evolution models of Jupiter and Saturn have shown, planets with deep radiative zones cool significantly faster compared to fully adiabatic cooling tracks. This affects not only the radius of the planet: the cooler interiors -- even for old irradiated giant exoplanets -- can lead to (increased) hydrogen-helium separation \citep{guillot_effect_1995}.}

The existence of such a deep radiative zone would significantly affect the evolution and therefore characterisation of giant exoplanets. As we enter the era of high-precision exoplanetary science with the \textit{James Webb} Space Telescope (JWST; \citealt{Gardner2006}) and the upcoming Ariel mission \citep{Tinetti2018}, characterising the interiors of warm Jupiters has become a primary objective. If radiative zones are a common feature of giant planets, as seems to be the case for Jupiter \citep{2023ApJ...952L..27B,2023NatAs...7..678C,aglyamov_alkali_2025,zhang_alkali_2026}, they could affect the cooling and act as a barrier to efficient mixing of the atmosphere with the interior. This raises critical questions: Can we use the measured atmospheric metallicity observed by transit spectroscopy to determine the planetary bulk composition? How do such deep radiative zones affect the observable radii of exoplanets over gigayear timescales {and for different equilibrium temperatures?} 

In this paper, we systematically investigate the role of deep radiative zones in the evolution of warm giant planets. Building on the mechanisms proposed for Jupiter, we explore how an opacity window near 2,000 K, where alkali metals could be depleted, influences the cooling and radius evolution of planets between 0.3 and {4.0 M$_{\rm{J}}$ with solar to ten-times solar envelope metallicities.} In Sect. \ref{sec:methods} we describe the thermal evolution models and the treatment of the opacity. We then apply these models to a set of {cold to} warm giant exoplanets in Sect. \ref{sec:results}, and assess the influence of opacity windows on planetary radii and interiors in Sect. \ref{sec:cooling_and_interior}. {We investigate in detail the dependence on the planetary mass, equilibrium temperature, and envelope metallicity in Sects. \ref{sec:t_eq} and \ref{sec:z_env}}. To illustrate {the effect of an opacity window on the characterisation of giant exoplanets}, we model the thermal evolution of HATS-49 b in Sect. \ref{sec:hats49b}. Our conclusions are presented in Sect. \ref{sec:conclusions}. Finally, additional results are presented in the appendix: In Appendix \ref{sec:appendix_heuristic} we derive a simple heuristic of how to scale the radii of giant planets with the bulk metallicity. {Further evolution models are presented in \editthree{Appendices \ref{sec:appendix_evolution}, \ref{sec:appendix_teq} and \ref{sec:appendix_zenv}}.}

\section{Methods}\label{sec:methods}

To model the evolution of irradiated giant planets, we used an early pre-release version of the Modules in Experiments in Stellar and Planetary Astrophysics (\textsc{MESPA}) code (\citealt{2025A&A...704A.253H}). This code implemented crucial modifications to the Modules for Experiments in Stellar Astrophysics (\textsc{MESA}; \citealt{Paxton2011,Paxton2013,Paxton2015,Paxton2018,Paxton2019,2023ApJS..265...15J}) stellar evolution code, making it better suitable for giant planets. To model the thermal evolution, the equations of planetary structure and evolution \citep[e.g.][]{Kippenhahn2012} were solved with the Henyey method \citep{Henyey1965}. 

The \textsc{MESPA} code replaces the equations in state of \textsc{MESA} and implements various mixtures of hydrogen, helium and heavy elements. In this work we used the hydrogen-helium equation of state from \citet{2021ApJ...917....4C}, and a 50-50 (by mass) water-rock mixture from \editthree{the Quotidian equation of state (QEOS; \citet{More1988,Vazan2013}).} The mixtures were calculated with the ideal mixing approximation (see \citealt{Mueller2020} and \citealt{2025A&A...704A.253H} for details).

A crucial ingredient in the cooling of giant planets is the determination of whether convection or radiation and conduction transport the heat out of the planet. The models in this work did not assume a specific energy transport mechanisms, instead the dominant mode of transporting energy was determined locally by the Schwarzschild criterion \citep{schwarzschild_equilibrium_1906}. Notably the radiative temperature gradient is proportional to the opacity. Therefore, the dominant mode of energy transport is influenced by the exact value of the opacity, and can significantly influence the cooling of the planet by dictating whether convection develops and how efficiently heat is transported in radiative layers. 

Early opacity calculations for giant planets suggested that there is a transparency window of hydrogen at temperatures of $\sim$ 2,000 K that would shut down convection \citep{1994Icar..112..354G,1994Icar..112..337G}. Later calculations, however, showed that whether this opacity window is enough to shut down convection strongly depends on the amount of alkali metals that are present \citep{Guillot2004,Freedman2008,siebenaler_conditions_2025}. Notably, there is evidence that alkali metals are in fact strongly depleted on Jupiter  \citep{2023ApJ...952L..27B} and that their abundance could deviate significantly from solar values. To account for this potential depletion, we followed previous work and implemented a parametrised reduction of the radiative opacity at temperatures of $\sim$ 2,000 K as follows \citep{muller_can_2024,2025A&A...704A.253H}:

\begin{equation}
    \kappa_r (\rho, T) = \kappa_0 (\rho, T) \left(1 - w \, \textrm{e}{^{-0.5 (\log T - \log T_0) / \Delta \log T}}\right) \, .
    \label{eq:opacity_scaling}
\end{equation}

\noindent
Here, $\kappa_r$ is the radiative opacity used in the evolution calculation, {$\kappa_0$ is the unmodified radiative opacity dependant on the local gas density $\rho$ and temperature $T$}, $\log T$ is the logarithm of $T$ in K, and $\log T_0$, $\Delta \log T$ and $w$ are free parameters. {For the low-temperature regime relevant for this work, we used tables privately communicated to the \textsc{MESA} developers by R.~S.~Freedman in 2011 (\textsc{kap\_lowT\_prefix = \enquote{lowT\_Freedman11}}), which were later published in \citet{Freedman2014}. For densities beyond the range of these tables, we used the standard approach in \textsc{MESA}, which is to set $\kappa(\rho > \rho_\textrm{max}, T) = \kappa(\rho = \rho_\textrm{max}, T)$}.
{The \citet{Freedman2014} tables top out at a pressure of 300 bar, which is lower compared to the pressures at which deep radiative zones commonly appear (see Sect. \ref{sec:results}). Beyond that, the opacities were extrapolated as described above. More recent opacity calculations suggest that opacity keeps increasing with pressure\footnote{{Increasing the pressure from 300 to 1000 bar at the same temperature appears to increase the opacity by about a factor of two \citep{siebenaler_conditions_2025}.}}, although there still remains a significant uncertainty for the conditions relevant in this work \citep{siebenaler_conditions_2025}.}

To match the location and width of the opacity window present when alkali metals are depleted \citep{1994Icar..112..354G, Freedman2008,siebenaler_conditions_2025}, we set $\log T_0 = 3.3$ and $\Delta \log T = 0.15$. The opacity factor $w$ parametrises the extent of the reduction {and can be interpreted as the amount of alkali-metal depletion.} A value of $w = 0.90$ results in a similar reduction as in \cite{1994Icar..112..354G}, and would correspond to a depletion of alkali metals of about $10^{-4}$ relative to solar abundances \citep{siebenaler_conditions_2025}.

For the outer boundary condition we adopted the commonly used irradiated grey approximation of \citet{Guillot2010,Guillot2011}. Two free parameters in that model are the surface pressure, which we set to $P_{\rm{surf}} = 0.1$ bar, and the $\gamma$ parameter, which is the ratio of the visible and infrared opacities. To account for $\gamma$ being a function of the planetary equilibrium temperature, we used an empirical scaling law \citep{poser_effect_2024}.

As a starting point for the evolution models we assumed a hot start, where the planet forms inflated and with high entropy \citep[e.g.][]{2012ApJ...745..174S}. We also assumed that the planets were fully formed and that the protoplanetary disk has disappeared. For simplicity, all models used a 10 M$_\oplus$ heavy-element core with a sharp boundary, and a {protosolar hydrogen-helium ratio}.

\section{Results}\label{sec:results}

In this section we present the outcome of thermal evolution models for various planetary masses, equilibrium temperatures, {envelope metallicities}, and opacity scaling factors. For the masses and equilibrium temperatures we considered the ranges of {cold to} warm Jupiters, namely {$M = 0.3$ to $4.0$ M$_{\rm{J}}$, $Z_{\rm{env}} = 1$ to $10$ Z$_\odot$, and T$_{\rm{eq}} = 100, 125, 150, 175, 200, 400$ and $800$ K. To investigate how the depth of the opacity window effects the results, {we used $w = 0.00, 0.50$, and $0.90$.} We first give an overview of the different planetary structures present in our evolution models, and how they affect the cooling. We then specifically discuss the influence of the planetary equilibrium temperature and envelope metallicity, before ending this section with a concrete example of how the inferred planetary bulk metallicity could be affected if there is an opacity window.}

\subsection{Overview of the different outcomes}\label{sec:overview}

{We generally observed three different structures in our evolution models:}

\begin{enumerate}
    \item {A thin radiative region at low pressures ($\lesssim$ 1 bar) on top of a large convective envelope. This is common for low planetary equilibrium temperatures ($\lesssim 150$ K) and younger planetary ages.}
    \item {A more complicated structure characterised by several distinctive layers. Starting from the top, there is a shallow radiative zone down to $\sim$ 10 to 100 bar on top of a fairly narrow convection zone ending at $\sim 10^3$ bar. Between $\sim 10^3$ and $\sim 10^4$ bar, there is a second detached deep radiative zone transitioning into a deep convective envelope.}
    \item {A single thick radiative zone down to pressures of $\gtrsim 10^3$ bar on top of a convective envelope. This is the common outcome for high equilibrium temperatures ($\gtrsim 600$ K).}
\end{enumerate}

\begin{figure*}
    \centering
    \includegraphics[width=\linewidth]{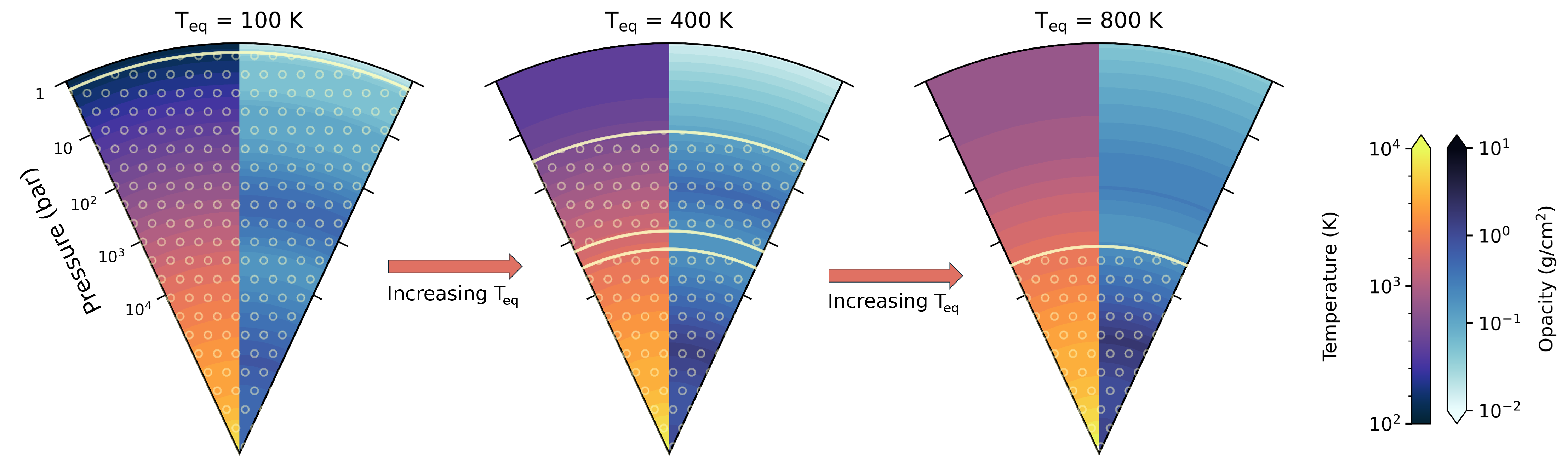}
    \caption{{Schematic illustrations of the three planetary structures predicted by our models. Convective boundaries and regions are marked with lines and circles. The three slices show the structures at 7 Gyr for different equilibrium temperatures, $T_{\rm{eq}} = 100, 400,$ and $800$ K, and otherwise identical model parameters. {The intrinsic temperatures at the outermost radiative convective boundaries are about 82, 92, and 97 K, respectively.} The left and right sides of the slices are coloured by the local temperature and opacity (see the colour bars). For low $T_{\rm{eq}}$, the planet is fully convective, with a thin radiative photosphere at low pressures (below about 1 bar). At intermediate $T_{\rm{eq}}$, there is a thicker upper radiative zone on top of a convective zone with a second detached deep radiative zone underneath. At high $T_{\rm{eq}}$, a single continuous deep radiative zone is present. Note that the transition values of $T_{\rm{eq}}$ between the three structures depend on the planetary mass, age, envelope metallicity, and opacity factor.}}
    \label{fig:overview}
\end{figure*}

\noindent
{These different structures are sketched in Fig. \ref{fig:overview} where we also show how the temperature and opacity varies with depth. These particular models were constructed for a 1 M$_{\rm{J}}$ planet with a solar metallicity envelope, $w = 0$ and $T_{\rm{eq}} = 100, 400,$ and $800$ K, and correspond to a planetary age of 7 Gyr. {At this age, the intrinsic temperatures at the outermost radiative convective boundaries were approximately 82, 92, and 97 K for the three models (from lowest to highest $T_{\rm{eq}}$).} We emphasise that the values of the equilibrium temperature at which one structure morphs into the other is affected by the planetary mass, envelope metallicity, age, and the opacity factor, and should therefore be seen as rough estimates.}

{As demonstrated in previous studies \citep[e.g.][]{burrows_radii_2000, arras_thermal_2006, thorngren_intrinsic_2019}, the general trend is that with increasing equilibrium temperature, the radiative-convective boundary moves inwards, since the strong irradiation flattens the temperature gradient in the envelope. As can be clearly seen in Fig. \ref{fig:overview}, the opacity does not monotonically increase deeper inside the planet, but rather there is a local minimum at $\sim 10^3$ to $10^4$ bar. Under certain conditions, the opacity there is low enough that a second, deep radiative zone can form that is sandwiched between two convective regions. In the upcoming sections, we explore how the cooling of giant planets depends on the various model parameters, and also the conditions for which deep radiative zones can form.}

{As previously noted in \citet{muller_can_2024}, the region up to $10^3$ bar contains very little mass: For a 1 M$_{\rm{J}}$ planet, the outer radiative and convective zones contain only $\sim 10^{-4}$ M$_{\rm{J}}$ ($\sim 10^{-2}$ M$_\oplus$). Consequently, material that is accreted as the planet cools, for example asteroids or comets, can change the composition of this outer region. Our results imply that for irradiated warm giant exoplanets, the presence of a deep radiative zone could be the norm rather than the exception. This directly affects the interpretation of atmospheric composition measurements, for example by JWST observations or the upcoming Ariel mission. We argue that atmospheric composition measurements should not be taken as an upper bound for the planetary bulk composition. The atmospheric composition also likely does not represent the bulk composition, if giant exoplanets have inhomogeneous interiors similar to those of Jupiter and Saturn \citep[e.g.][]{Debras2019,mankovich_diffuse_2021,Helled2024}.}

\subsection{Effect on the cooling, structure, and characterisation}\label{sec:cooling_and_interior}

{We first present results for a few specific example parameters, before diving deeper into how the planetary mass, equilibrium temperature, and envelope metallicity affects the thermal evolution.} 

\begin{figure}
    \centering
    \includegraphics[width=\linewidth]{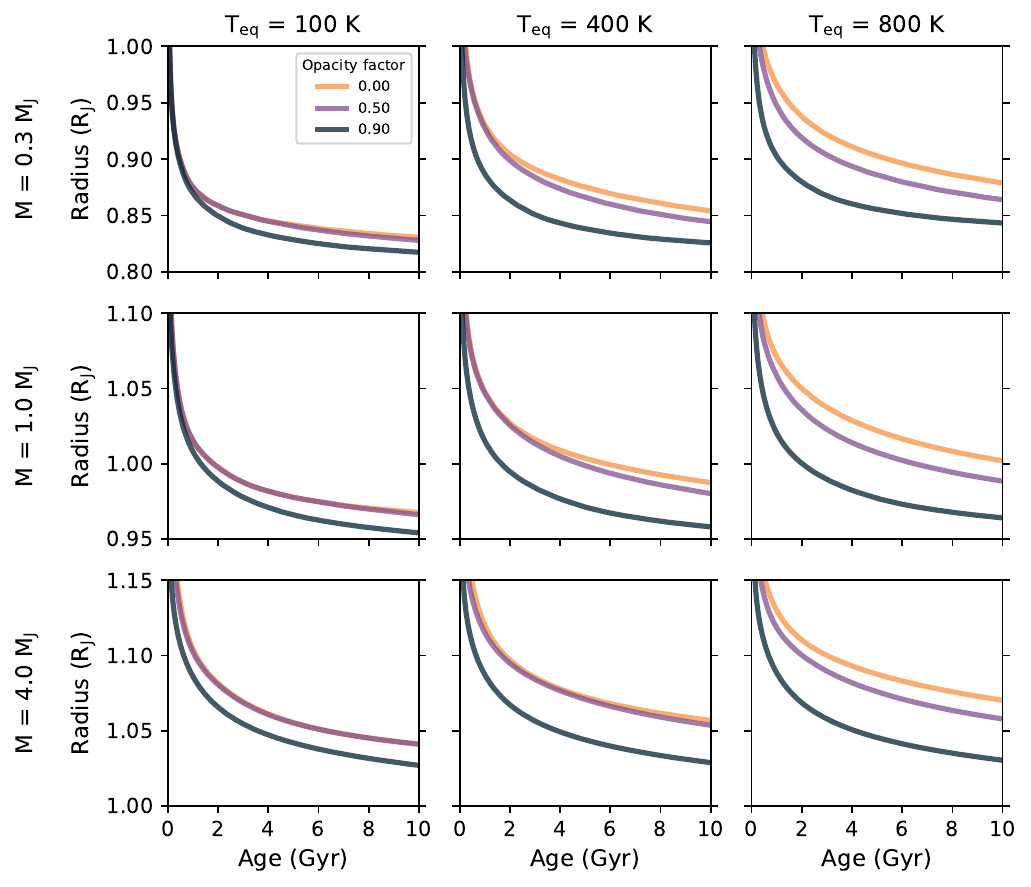}
    \caption{Radius evolution for different planetary masses (\textit{rows}), equilibrium temperatures (\textit{columns}), opacity factors (coloured lines; see the legend), {and a solar-metallicity envelope.} The radius difference between the various opacity factors increases with the equilibrium temperature and can reach a relative difference of up to about 5\%.}
    \label{fig:radius_evolution}
\end{figure}

Figure \ref{fig:radius_evolution} shows how the cooling and sizes of planets change depending {on the planetary mass and how deep the opacity window is (different values of $w$) for models with a solar-metallicity envelope (see Appendix \ref{sec:appendix_evolution} for different envelope metallicities).} {At the coldest equilibrium temperature we considered (100 K), the opacity has to be significantly depleted ($w = 0.90$) for the cooling to be affected. This is because the deep radiative zone either does not appear at all, or only once the planet is very old (see Sect. \ref{sec:t_eq}).} In all the {other} cases the cooling was strongly affected. Increasing the depth of the window causes planets to cool faster, leading to a smaller planet at any given time. Our calculations show that the effect size increases with both $M$ and $T_{\rm{eq}}$, with the largest differences reaching a relative difference of 5\% in radius. {Additional models with super-solar metallicity envelopes (up to 10 times) show that this trend generally continues; however, the effect is sometimes weaker because the enhanced opacity can delay or stop the deep radiative zone from appearing (see Sect. \ref{sec:t_eq}).} Considering that current observational uncertainties of giant exoplanet radii are {commonly around a few percent}, this is a non-negligible difference. 

{A lower opacity could} also significantly change the estimated bulk metallicity, which is commonly inferred by using evolution models that match the observed planetary radius \citep[][]{Fortney2010,Thorngren2016,Muller2023}. As we discuss in Appendix \ref{sec:appendix_heuristic}, a 5\% change in radius changes the inferred bulk metallicity by about 10 percentage points (in absolute terms; for example, from 30\% to 40\%). In Sect. \ref{sec:hats49b} we demonstrate this for the exoplanet HATS-49 b.

An opacity window can also create deep radiative zones. {Although deep radiative zones and their effect on the thermal evolution were previously discussed for Jupiter and Saturn \citep{1994Icar..112..354G,1994Icar..112..337G,guillot_effect_1995,2023A&A...680L...2H,muller_can_2024}}, they are generally not considered in the context of giant exoplanets. Beyond affecting the cooling of a giant planet, deep radiative zones can also `disconnect' the deeper interior from the atmosphere in terms of its composition. This is because material transport through such radiative zones is potentially very slow, which can result in the atmospheric composition being very different from the bulk of the planet. Previous simulations even suggest that the deep radiative zone is stable enough to support an atmospheric metallicity that is, in fact, higher than deeper inside the planet \citep{muller_can_2024}. 
{The requirement for this enrichment to remain over the course of a planet's thermal evolution (a few gigayears) is that the vertical transport of material has to be slow enough. The diffusion coefficients would have to be smaller than about $10^{-2}$ cm$^2$/s, unless the enrichment is more recent \citep{muller_can_2024}.}

\begin{figure}
    \centering
    \includegraphics[width=\linewidth]{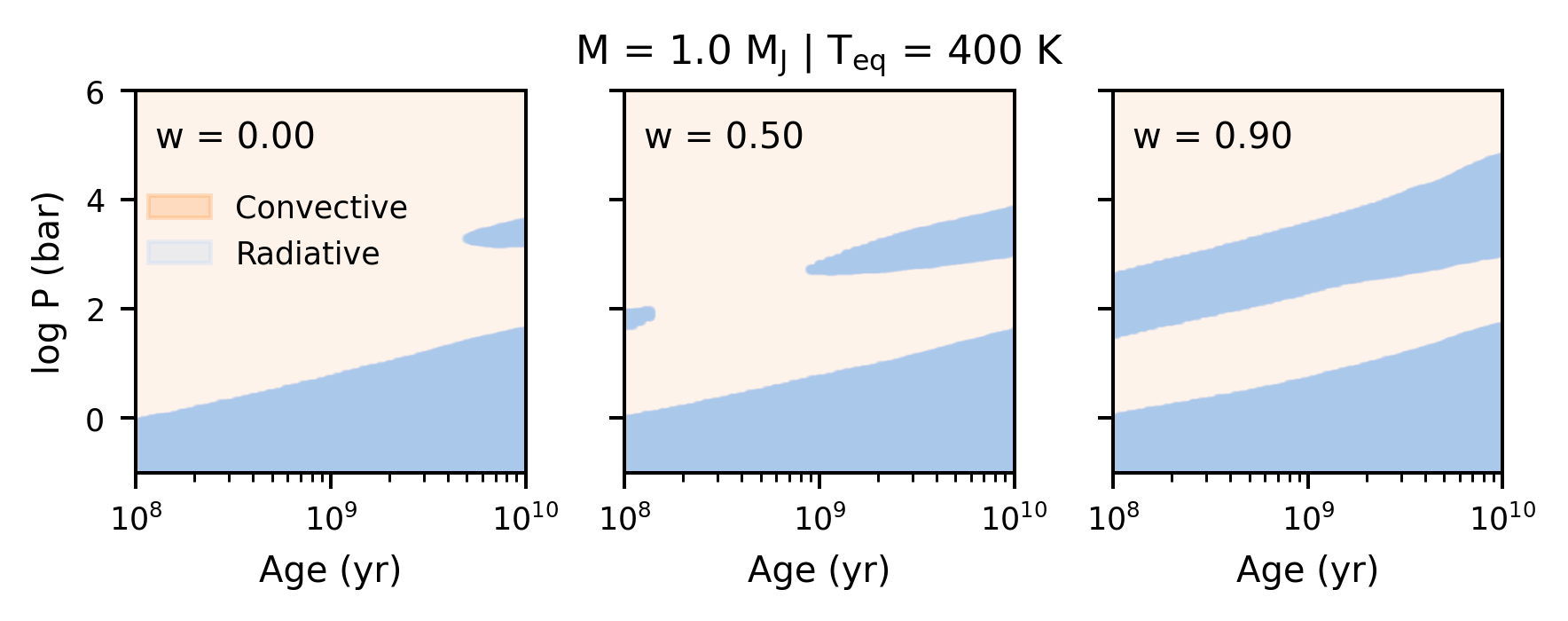}
    \caption{Kippenhahn diagrams showing the evolution of convective (orange) and radiative zones (blue) with time for $M = 1.0 \,\rm{M}_{\rm{J}}$, $T_{\rm{eq}}$ = 400 K, {a solar-metallicity envelope}, and $w = 0, 0.50$, and $0.90$. In all cases the planets develop two radiative zones: a first one at the planetary surface, and a second deep radiative zone at higher pressures whose thickness and time of appearance depends on the opacity.}
    \label{fig:kippenhahn_m1t400}
\end{figure}

Figure \ref{fig:kippenhahn_m1t400} shows how radiative and convective regions appear and evolve for an example planet ($M =$ 1 M$_{\rm{J}}$ planet, $T_{\rm{eq}} = 400$ K, {solar-metallicity envelope}) for three different values of $w = 0.00, 0.50$, and $0.90$. Remarkably, {in this particular case}, after at most 4 Gyr deep radiative zones are present in all the simulations, even if the opacity was unmodified ($w = 0$). {This has not been previously shown, and is different compared to evolution models of Jupiter that required an ad hoc opacity reduction for the deep radiative zone to appear at an age of about 4.5 Gyr \citep{2023A&A...680L...2H,muller_can_2024}.} When the deep radiative zone appears and how thick it is depends crucially on the depth of the opacity window, equilibrium temperature, {envelope metallicity, and the planetary mass, as will be shown in Sects. \ref{sec:t_eq} and \ref{sec:z_env}}. The top of the deep radiative zone generally is at a pressure of about $10^3$ bar, and reaches down towards one or a few times $10^4$ bar depending on $w$ and the planetary age. Both the top and the bottom of the deep radiative zone move inwards as the planet cools.

\begin{figure}
    \centering
    \includegraphics[width=0.75\linewidth]{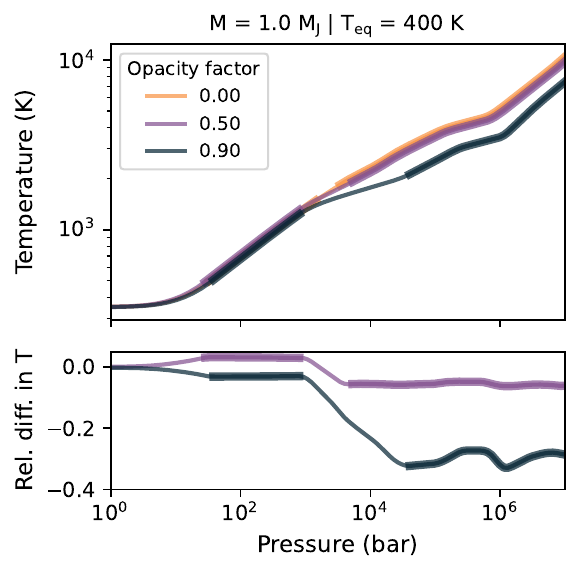}
    \caption{{Temperature profiles and their differences for three opacity factors.} \textit{Top}: {Pressure-temperature profile at 7 Gyr for a planetary mass of 1 $\rm{M}_{\rm{J}}$, solar envelope metallicity, equilibrium temperature of 400 K, and three different opacity factors, $w = 0, 0.50,$ and $0.90$.} The lines are coloured according to $w$, and convective regions are marked with thicker lines. \textit{Bottom}: Difference in the logarithmic temperature as a function of pressure. Deep opacity windows lead to significantly colder deep interiors, which relative differences of up to about 35\%.}
    \label{fig:logP_logT_interior}
\end{figure}

Figure \ref{fig:logP_logT_interior} shows the corresponding pressure-temperature profiles at {an advanced planetary age of 7 Gyr}. Again, the depth and thickness of the deep radiative zones and how they depend on $w$ are clearly visible. These interior profiles also demonstrate that a deep opacity window leads to faster cooling and significantly colder interiors (as expected from the smaller predicted sizes). For the largest opacity factor ($w = 0.90$), interiors were up to about 30\% colder compared to the baseline. This explains the large reduction in radius that we previously observed. {It is interesting to consider that the significantly faster cooling and therefore cooler interiors could also cause helium rain even for irradiated giant exoplanets. However, the current uncertainties in the hydrogen-helium phase diagram are too larger to draw robust conclusions on this possibility \citep{mankovich_bayesian_2016, 2020ApJ...889...51M,howard_evolution_2024, sur_evolution_2025,sur_next-generation_2026}.}

\subsection{Dependence on the equilibrium temperature}\label{sec:t_eq}

{As already hinted at in the previous two sections, the equilibrium temperature plays an important role in determining the number and depth of the deep radiative zones in planetary envelopes. To investigate this dependence further, we calculated evolution models for $M = 0.3, 1.0,$ and $4.0$ M$_{\rm{J}}$, $T_{\rm{eq}} = 100, 150, 200, 400, 600,$ and $800$ K for varying $w = 0, 0.50,$ and $0.90$. The metallicity in the envelope was solar, and additional models with $M = 0.6$, $1.5,$ and $2.0$ M$_{\rm{J}}$ are presented in Appendix \ref{sec:appendix_teq}.}

\begin{figure*}
    \begin{subfigure}[b]{\linewidth}
        \centering
        \includegraphics[width=0.75\linewidth]{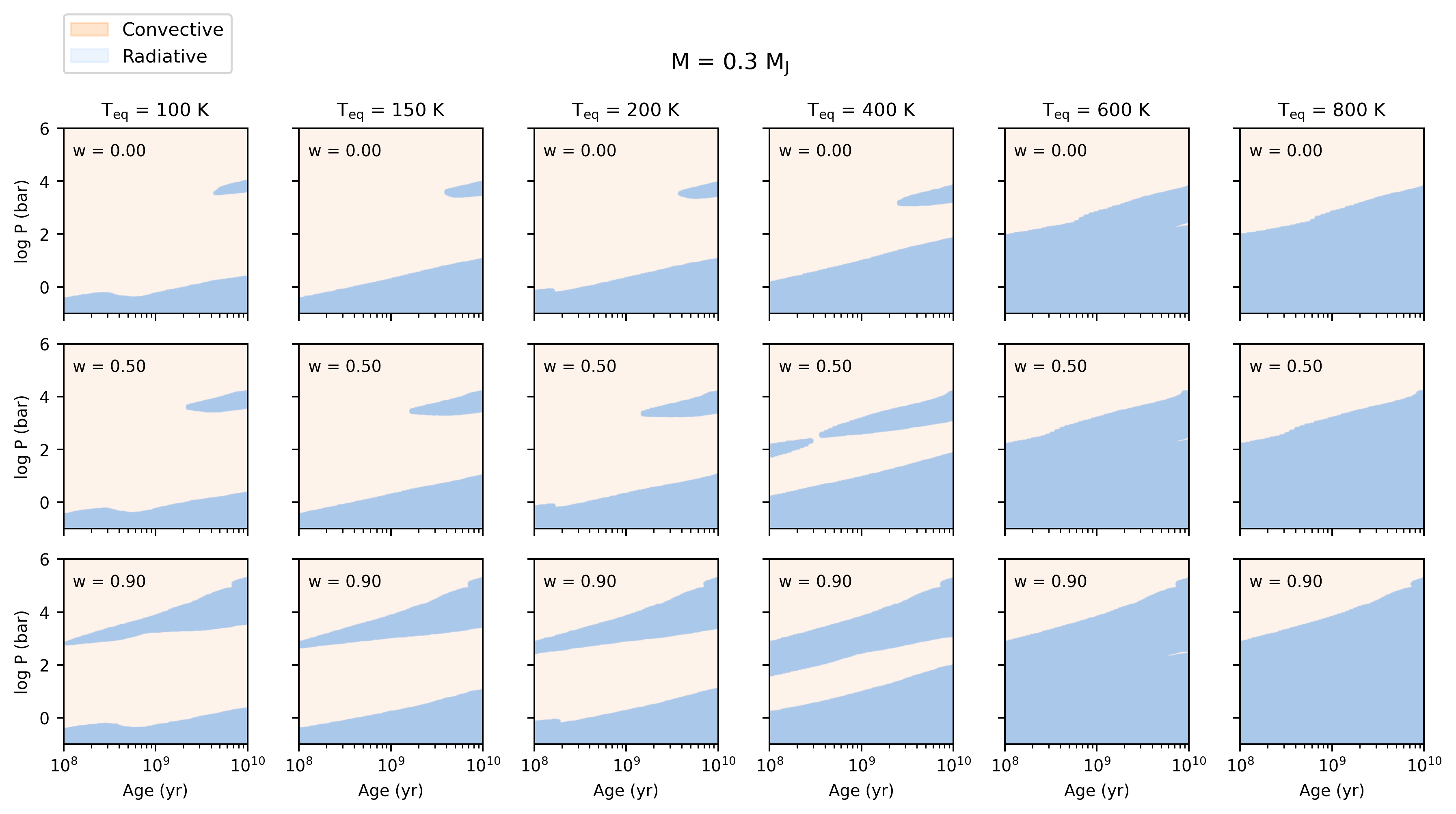}
    \end{subfigure}
    \begin{subfigure}[b]{\linewidth}
        \centering
        \includegraphics[width=0.75\linewidth]{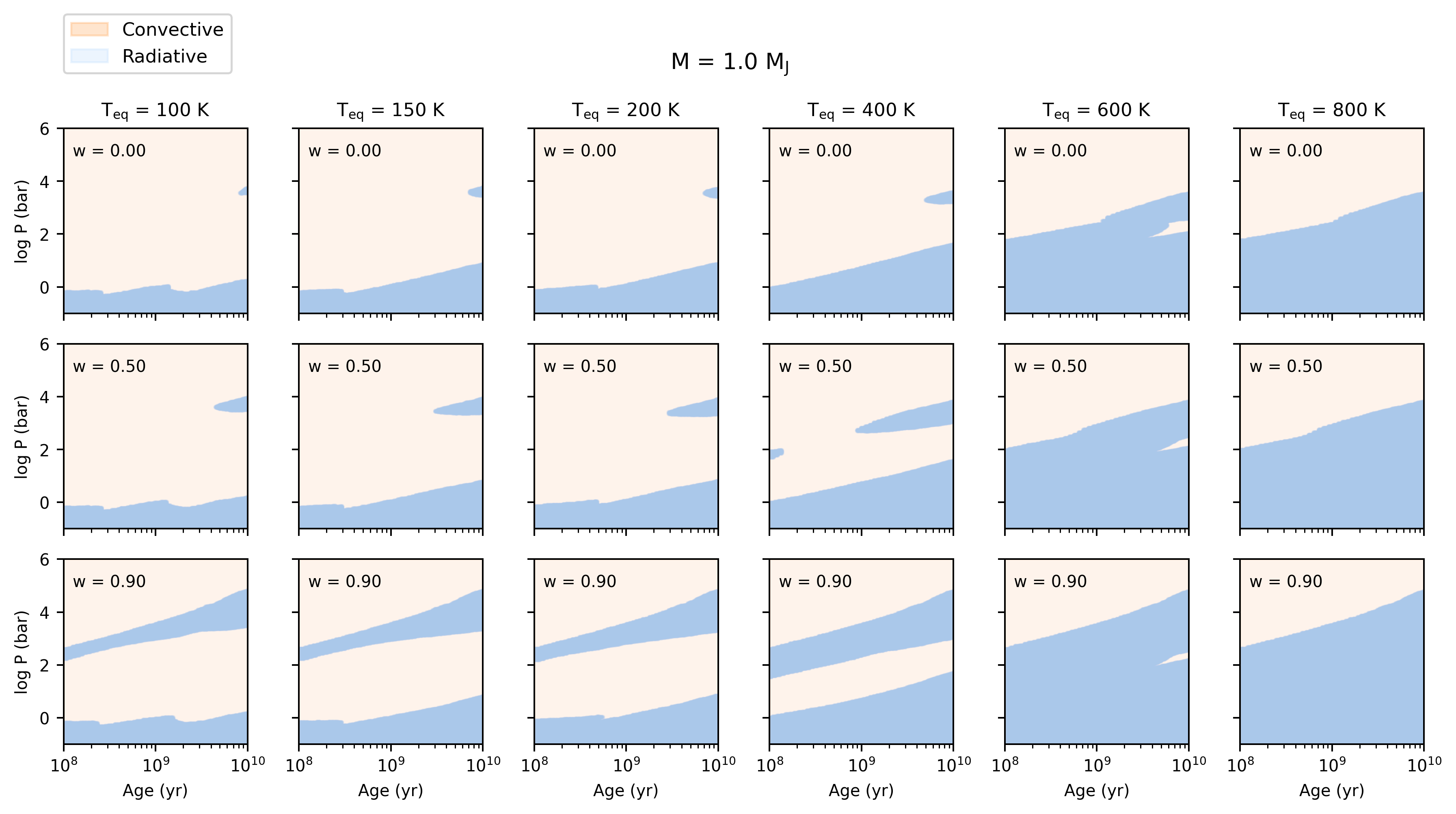}
    \end{subfigure}
    \begin{subfigure}[b]{\linewidth}
        \centering
        \includegraphics[width=0.75\linewidth]{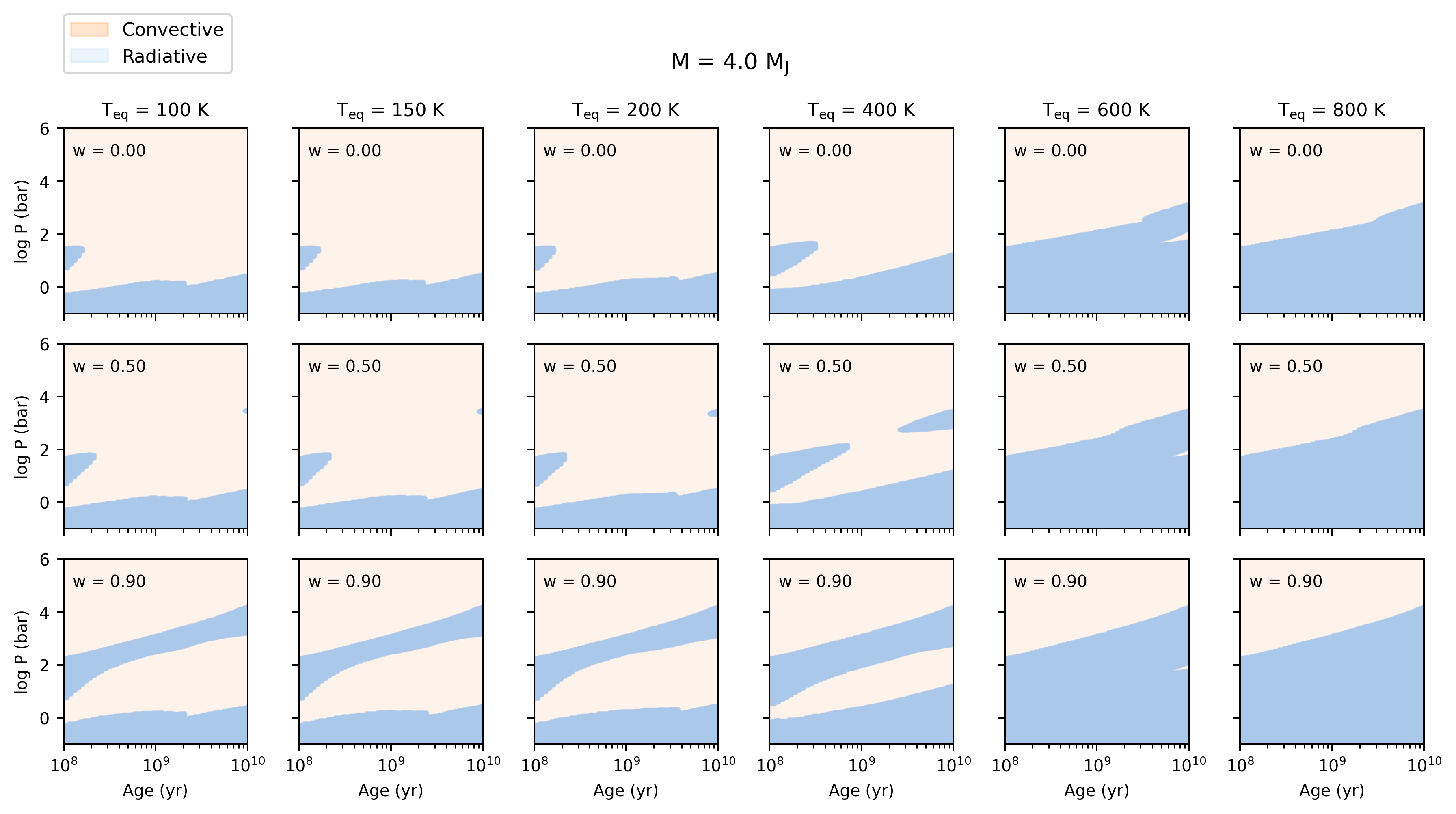}
    \end{subfigure}
    \caption{{Kippenhahn diagrams showing the evolution of convective (orange) and radiative zones (blue) with time for planetary masses of $M = 0.3$ (\textit{top}), $1.0$ (\textit{centre}), and $4.0$ $\rm{M}_{\rm{J}}$ (\textit{bottom}). The columns and rows correspond to different equilibrium temperatures and opacity factors.}}
    \label{fig:kippenhahn_teq}
\end{figure*}

{The results are shown in Fig. \ref{fig:kippenhahn_teq}. The first obvious trend which we already discussed in the previous section is that the radiative zone extends to progressively higher pressures as the stellar irradiation increases. If, when and at which $T_{\rm{eq}}$ a detached deep radiative zone appears clearly also depends on the planetary mass and the value of $w$. With unmodified opacities ($w = 0$) in the low-mass model ($M = 0.3$ M$_{\rm{J}}$), a detached deep radiative zone eventually develops for equilibrium temperatures between 100 and 600 K; at 800 K, there is a single large radiative zone at least from 10 Myr onwards. For $M =  1.0$ M$_{\rm{J}}$ the behaviour is similar, although the detached deep radiative zones appear later, and once again merge into one single deep radiative layer for $T_{\rm{eq}} = 800$ K. The high-mass model ($M = 4.0$ M$_{\rm{J}}$) exhibits a quite different behaviour: While detached deep radiative zones are present at early times ($\sim 10$ Myr), for all but the model with $T_{\rm{eq}} = 600$ K they are transient and disappear before 1 Gyr.}

{A significant reduction of the opacity ($w = 0.90$) produces strikingly different results: A detached deep radiative zone is present for all masses up to $T_{\rm{eq}} = 800$ K from at least 10 Myr to 10 Gyr. For an intermediate reduction in the opacity $w = 0.50$ the results are somewhere between the previous two cases. While most models for $w = 0.50$ and $0.90$ eventually have a similar radiative-convective layering at 10 Gyr, the appearance of the deep radiative zone is generally delayed, which influences the cooling significantly (see also Fig. \ref{fig:radius_evolution} and Appendix \ref{sec:appendix_evolution}).}

{The general trend deduced from these models is that detached deep radiative zones develop earlier and at lower equilibrium temperatures ($T_{\rm{eq}} = 100$ to $600$ K) for low masses, and (if it all) later and at higher equilibrium temperatures for higher masses. As already noted in the previous section, this is true even for unmodified opacities ($w = 0$). Without reducing the opacity significantly, the deep radiative zones generally appear rather late, and therefore the cooling is influenced less. Also, in that case there is a reduced potential to enrich the atmosphere, because the deeper interior stays connected via large-scale convection to the atmosphere for most of the evolution.}

\subsection{Dependence on the envelope metallicity}\label{sec:z_env}

\begin{figure*}
    \centering
    \begin{subfigure}{0.49\linewidth}
        \centering
        \includegraphics[width=\linewidth]{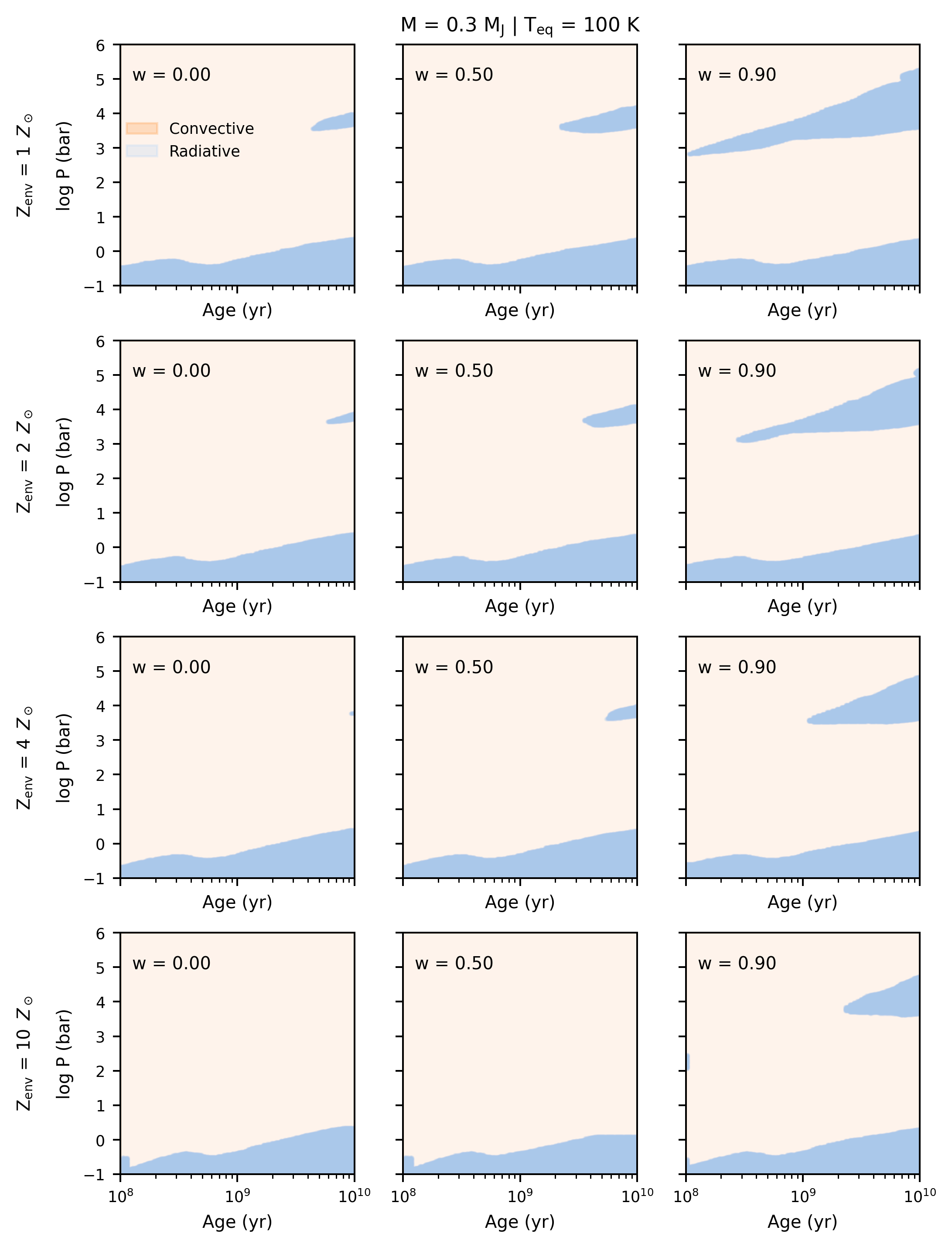}
    \end{subfigure}
    \begin{subfigure}{0.49\linewidth}
        \centering
        \includegraphics[width=\linewidth]{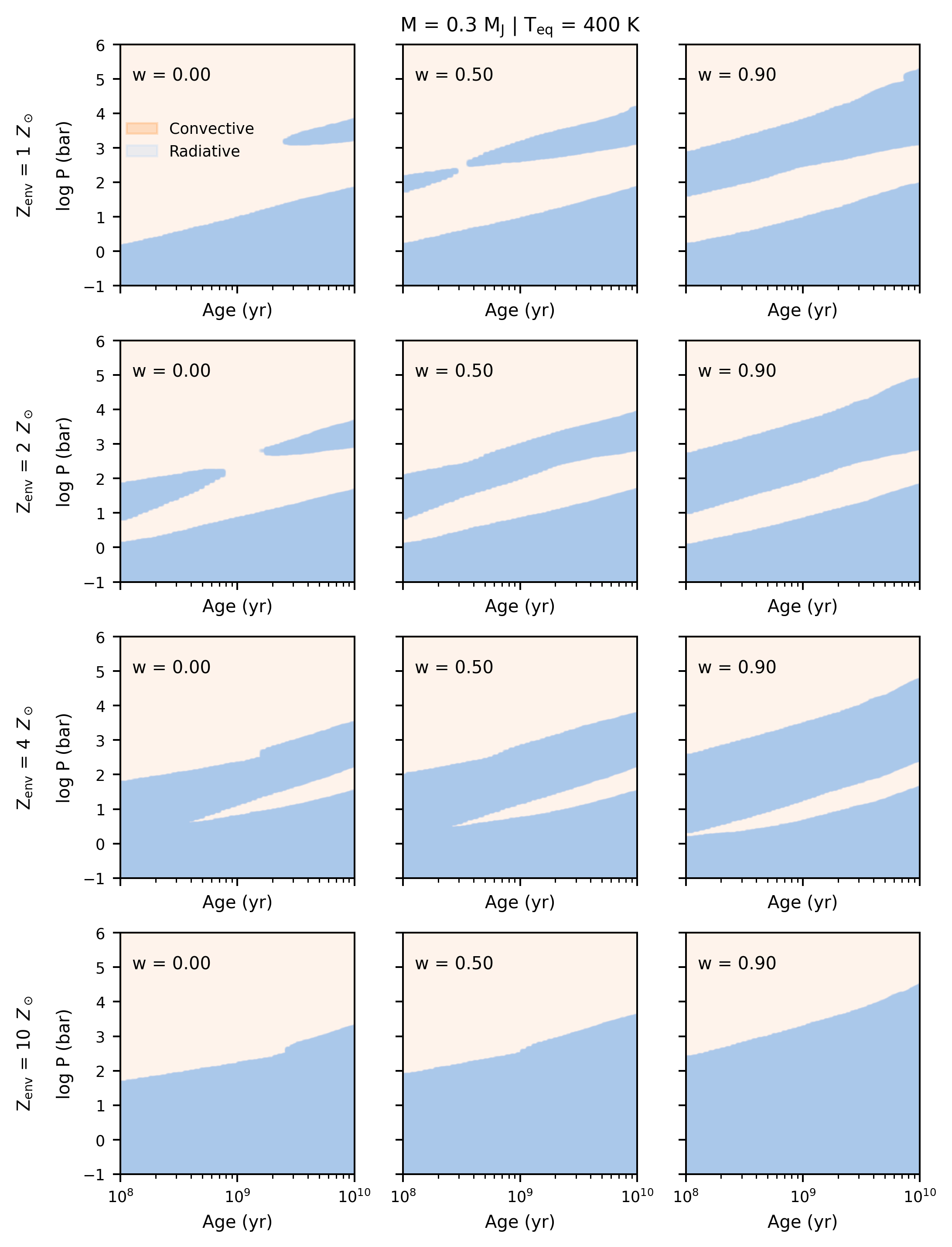}
    \end{subfigure}
    \caption{{Kippenhahn diagrams showing the evolution of convective (orange) and radiative zones (blue) with time for a planetary masses of $M = 0.3$ M$_{\rm{J}}$ for $T_{\rm{eq}} = 100$ (\textit{left}) and $400$ K (\textit{right}). The rows and columns correspond to different envelope metallicities and opacity factors.}}
    \label{fig:kippenhahn_zenv_03}
\end{figure*}

{All the models above assumed a solar-metallicity envelope -- an obvious simplification, considering the large varieties of atmospheric metallicities even in the solar system. Here, we present additional models where we varied the envelope metallicity from 1 to 10 times solar (in terms of mass). The results are shown in Figs. \ref{fig:kippenhahn_zenv_03}, \ref{fig:kippenhahn_zenv_10}, and \ref{fig:kippenhahn_zenv_40} for planetary masses of $0.3$, $1.0,$ and $4.0$ M$_{\rm{J}}$ and $T_{\rm{eq}} = 100$ and $400$ K. As before, we used $w = 0.00$, $0.50$, and $0.90$.}

{At $T_{\rm{eq}} = 100$ K, we find that an enriched envelope either delays or completely suppresses the arising of detached deep radiative zones. With the exception of the most massive model with the most enriched envelope, we find that for $w = 0.90$ deep radiative zones are always present at the end of the simulation, even though their arising is significantly delayed in many cases.}

{At warmer equilibrium temperatures ($T_{\rm{eq}} = 400$ K), we find a different behaviour: Detached deep radiative zones generally appear earlier and are not suppressed. In fact, for the $4.0$ M$_{\rm{J}}$ model the only one with a deep radiative zone at 10 Gyr has the highest envelope metallicity. Beyond affecting the shape and time of appearance, the envelope metallicity also affects at which pressures the deep radiative zones appear. Higher envelope metallicities pull the detached deep radiative zones to lower pressures (closer to the surface). Their bottoms are commonly around $10^3$ bar for a 10 times solar envelope metallicity. For example, for the case of 1.0 M$_{\rm{J}}$ and $w = 0$, at the end of the simulation, the deep radiative zone extends between about $10$ and $10^3$ bar. }

\begin{figure}
    \centering
    \begin{subfigure}{0.49\linewidth}
        \centering
        \includegraphics[width=\linewidth]{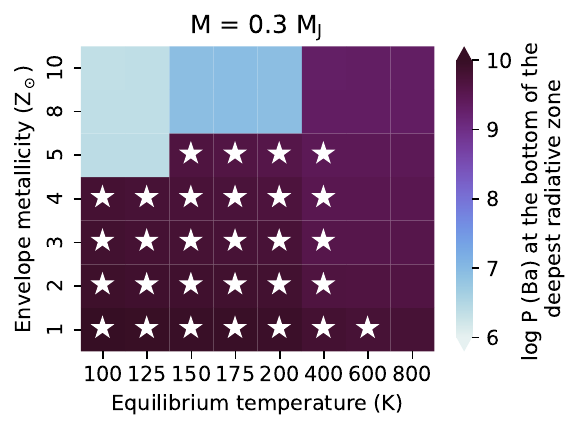}
    \end{subfigure}
    \begin{subfigure}{0.49\linewidth}
        \centering
        \includegraphics[width=\linewidth]{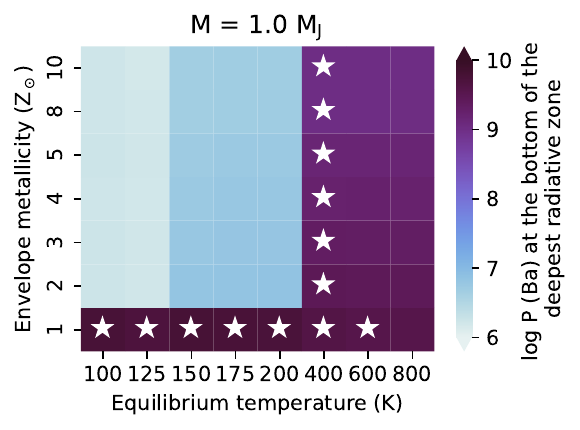}
    \end{subfigure}
    \begin{subfigure}{0.49\linewidth}
        \centering
        \includegraphics[width=\linewidth]{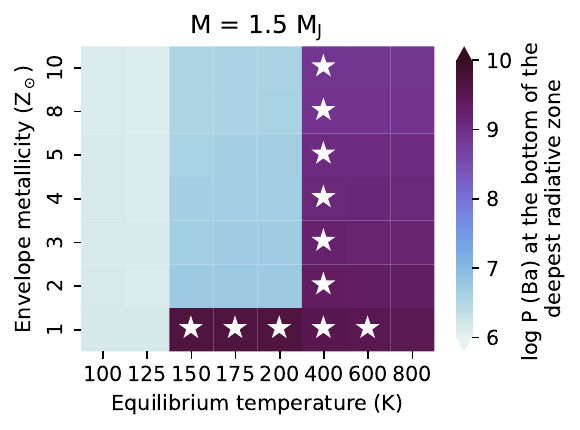}
    \end{subfigure}
    \begin{subfigure}{0.49\linewidth}
        \centering
        \includegraphics[width=\linewidth]{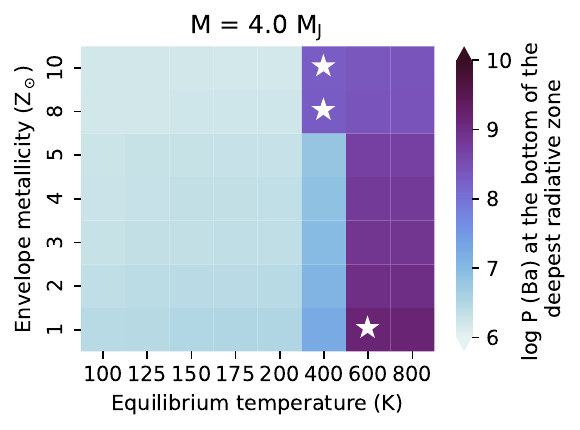}
    \end{subfigure}
    \caption{{Heat-maps showing the value of the pressure at the bottom of the deepest radiative zone as a function of the equilibrium temperature and envelope metallicity for planetary masses of $0.3$, $1.0$, $1.5,$ and $4.0$ M$_{\rm{J}}$. The interiors correspond to the final age of the model (10 Gyr) and with an unmodified opacity $(w = 0)$. The stars indicate the presence of detached deep radiative zones located between two convection zones (see Sect. \ref{sec:overview} and Fig. \ref{fig:overview}).}}
    \label{fig:hotbox}
\end{figure}

{Figure \ref{fig:hotbox} shows the combinations of planetary mass, equilibrium temperature, and envelope metallicity for which a deep radiative zone is present at 10 Gyr, together with the pressure at its lower boundary. These models used an unmodified opacity. For low-mass planets, only very enriched envelopes with $Z_{\rm{env}} \gtrsim 5 Z_\odot$ stop the deep radiative zones from appearing for $T_{\rm{eq}} = 100$ to $400$ K. The intermediate mass planets react differently: At $T_{\rm{eq}} \lesssim 400$ K, even a two-times solar enhancement is enough to completely suppress the deep radiative zones. However, at $T_{\rm{eq}} = 400$ K, they exist for all metallicities we considered. For the high-mass model, the deep radiative zone is only present for an extremely narrow set of parameters. At higher equilibrium temperatures, there is a single deep radiative zone.}

{In summary, we find that if the equilibrium temperature is low and the opacity is not reduced, an increased envelope metallicity suppresses deep radiative zones. This is no longer true for warm equilibrium temperatures, at which the effect is opposite and a higher metallicity accelerates the development of deep radiative zones and moves them towards shallower pressures. For strongly reduced opacities ($w = 0.90$), deep radiative zones exist for (almost) all combinations of masses, equilibrium temperatures, and envelope metallicities we considered. This affects their cooling, as discussed in Appendix \ref{sec:appendix_evolution}. For $T_{\rm{eq}} = 400$ K, they exist also at very young ages (10 Myr) and are stable until at least 10 Gyr. This configuration could allow for significant enrichment of the atmosphere with heavy elements.}

\subsection{Example: HATS-49 b}\label{sec:hats49b}

\begin{figure}
    \centering
    \includegraphics[width=0.9\linewidth]{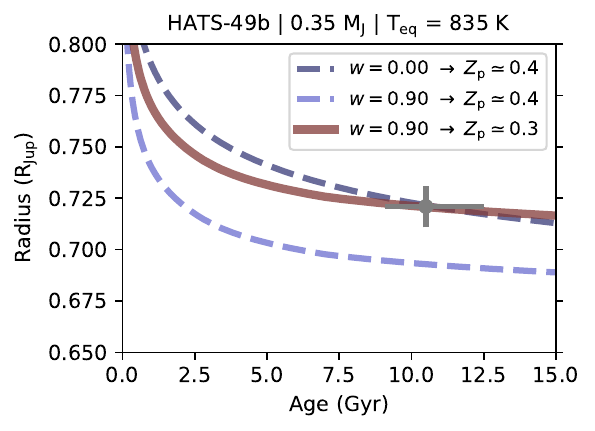}
    \caption{Radius as a function of time for different values of $w$ and bulk metallicities (see the legend). The radius and age of HATS-49 b are shown as grey error bars. To match the observations, HATS-49 b can have either 40\% ($w = 0$) or 30\% ($w = 0.9$) of heavy elements by mass, depending on the depth of the opacity window.}
    \label{fig:hats49b_radius}
\end{figure}

We modelled the thermal evolution of HATS-49 b to demonstrate the effect of a deep opacity window on the interior characterisation. The resulting cooling curves are shown in Fig. \ref{fig:hats49b_radius}. Using the observational data from \citet{hartman_hats-47b_2020}, we first inferred that the planet has a bulk metallicity of around $Z_{\rm{p}} \simeq 0.4$ if the opacity is unmodified ($w = 0$). Next, we calculated the thermal evolution using the same $Z_{\rm{p}}$ but with $w = 0.90$, implying that there is a deep dip in the opacity at a temperature of a few thousand kelvins. It is clear that this new cooling curve does not agree with the observed radius and age of HATS-49 b. We therefore ran further calculations and inferred that in this case, a bulk metallicity of $Z_{\rm{p}} = 0.3$ is required to match the observations.

These results demonstrate that the characterisation of a giant exoplanet is strongly affected by the modeller's choice of whether to reduce the opacity or not. In the case of HATS-49 b, the difference was about 10 percentage points (or 25\% in relative) terms. Compared to the formal uncertainty of the bulk metallicity as a result of standard errors of the planetary parameters \citep[e.g.][]{Thorngren2016,2023A&A...669A..24M,howard_giant_2025}, this adds a significant additional modelling uncertainty.

{For simplicity, we assumed that HATS-49 b has a solar-metallicity envelope and atmosphere. This neglects the possibility that there is a correlation between metal-rich atmospheres and interiors, which would significantly complicate the problem. Therefore, our results should not be interpreted as giving a robust estimate for the bulk metallicity of HATS-49 b. Rather, they illustrate that an opacity reduction at a temperature of a few thousand kelvins can significantly affect the planetary radius, and therefore also the inferred composition.}

\section{Conclusions}\label{sec:conclusions}

In this work we used thermal evolution models to investigate the influence of opacity windows and deep radiative zones on the cooling and characterisation of giant exoplanets. Our main findings are:

\begin{itemize}
    \item {Detached deep radiative zones arise as a generic outcome of giant planet thermal evolution rather than requiring an ad hoc opacity reduction. Even with unmodified radiative opacities, warm Jupiters older than $\sim$ 4 Gyr commonly develop a deep radiative zone. A reduction in the opacity causes these zones to appear earlier and over a broader range of planetary masses, equilibrium temperatures, and envelope metallicities.}
    \item {A deep opacity window accelerates the planetary cooling, yielding interiors that are colder by up to 35\% and radii that are smaller by up to 5\% relative to models with unmodified opacities.}
    \item {The planetary mass and equilibrium temperature jointly determine which of three characteristic interior structures develops (fully convective, layered with a single detached deep radiative zone, or a single thick radiative zone extending to depth \editthree{($\sim 10^4$ to $10^5$ bar)}, as well as the timing of the transition between these regimes. Deep radiative zones appear earliest and across a wide range of equilibrium temperatures for low-mass planets (0.3 to 1 M$_{\rm{J}}$), whereas for high-mass planets they are typically transient or require a substantially reduced opacity to persist.}
    \item {The envelope metallicity modulates the occurrence of deep radiative zones in a temperature-dependent manner: at low equilibrium temperatures ($T_{\rm{eq}} \sim 100$ K), an enriched envelope delays or suppresses the formation of a detached deep radiative zone, whereas at higher equilibrium temperatures ($T_{\rm{eq}} \gtrsim 400$ K) enrichment instead promotes their formation and shifts them to shallower pressures.}
 
\end{itemize}

{Our study shows that understanding opacities at high pressures is crucial for the modelling and characterisation of giant planets.} The potential existence of {detached deep radiative zones} suggests a profound `chemical decoupling' between a planet’s observable upper atmosphere and its deep interior. This implies that the metal enrichments measured by missions such as JWST and Ariel reflect localised envelope chemistry rather than the true bulk composition of the planet's gaseous envelope. {The opacities used in this work had to be extrapolated to higher pressures, which adds a significant amount of uncertainty regarding whether deep radiative zones exist naturally or require an additional reduction in the opacity, for example due to lower abundances of alkali metals. As we move towards a regime of high-precision exoplanet spectroscopy and detailed characterisation, accounting for the complex interplay of opacity windows and internal transport mechanisms is essential to bridge the gap between atmospheric observations and the fundamental physics of giant planet evolution and internal structure.

\begin{acknowledgements}
    We thank the anonymous referee for thoroughly reading our manuscript and providing useful comments. We acknowledge support from the Swiss National Science Foundation (SNSF) grant \texttt{\detokenize{200020_215634}} and the National Centre for Competence in Research ‘PlanetS’ supported by SNSF. This research used data from the NASA Exoplanet Archive, which is operated by the California Institute of Technology, under contract with the National Aeronautics and Space Administration under the Exoplanet Exploration Program. The thermal evolution models were calculated with a pre-release version of \textsc{MESPA} \citep{2025A&A...704A.253H}, which is based on the stellar evolution code \textsc{MESA} \citep{Paxton2011, Paxton2013, Paxton2015, Paxton2018, Paxton2019, jermyn_modules_2023}.
    Extensive use was also made of the \textsc{Python} packages \textsc{Jupyter} \citep{jupyter}, \textsc{Matplotlib} \citep{Hunter2007}, \textsc{NumPy} \citep{harris2020array}, \textsc{planetsynth} \citep{2021MNRAS.507.2094M}, \textsc{mesa\_reader} \citep{py_mesa_reader}, and \textsc{mesatools} \citep{pymesatools}.
\end{acknowledgements}

\bibliographystyle{aa}
\bibliography{library}

\begin{appendix}
\nolinenumbers  % remove line numbers for arXiv

\section{A simple heuristic for scaling planetary radii with the bulk metallicity}\label{sec:appendix_heuristic}

As the bulk metallicity of a giant planet increases, so does the mean density increases and therefore the planet gets smaller. While evolution models are required to accurately infer the bulk metallicity of giant planets, it can nonetheless be useful to have a simple approximation of the radius-metallicity interaction. Here, we briefly derive a simple heuristic of how to estimate the change in a planet's radius with the bulk metallicity. 

To demonstrate this scaling, we used the \textsc{planetsynth} evolution models \citep{2021MNRAS.507.2094M} and calculated the cooling of a 1 M$_{\rm{J}}$ planet with $T_{\rm{eq}} = 400$ K and bulk metallicities $Z_p = 0.05, ... ,0.40$. These cooling curves are shown in Fig. \ref{fig:appendix_heuristic} as the solid coloured lines. It is quite evident that the size difference between adjacent cooling curves appear regularly spaced. Indeed, calculating the mean differences over time shows that as the bulk metallicity increases by 5\%, the radius decreases by about half of that.

We calculated new cooling curves as follows: Let $R_{p, \, Z_i}$ be the radius of a planet with a metallicity $Z_i$. The radius of a planet with $Z_{i + 1} = Z_i + \Delta Z$ is then approximately $R_{p, Z_{i + 1}} = R_{p, \, Z_i} - 0.5 \, \Delta Z$, assuming $\Delta Z \geq 0$. These new evolution curves are shown in Fig. \ref{fig:appendix_heuristic} as the coloured dashed lines, which demonstrates that they approximate the real evolution curves rather well.

We therefore propose the following heuristic: A 1\% change in the radius of a giant planet can accommodate about a 2\% change in its bulk metallicity. We note, however, that the scaling clearly will depend on the assumed composition of the heavy elements (and other details of the model). The \textsc{planetsynth} models use a 50-50 water-rock mixture -- if, for example, the heavy elements were represented by water, one would expect to be able to change the metallicity more. {Additionally, for simplicity these models assumed a solar-metallicity envelope. If this would not have been the case, the situation becomes a lot more complicated, since metallicity-enhanced envelopes affect the opacity and therefore the cooling \citep[e.g.][]{Freedman2014, 2020ApJ...903..147M}.}

\begin{figure}[ht]
    \centering
    \includegraphics[width=\linewidth]{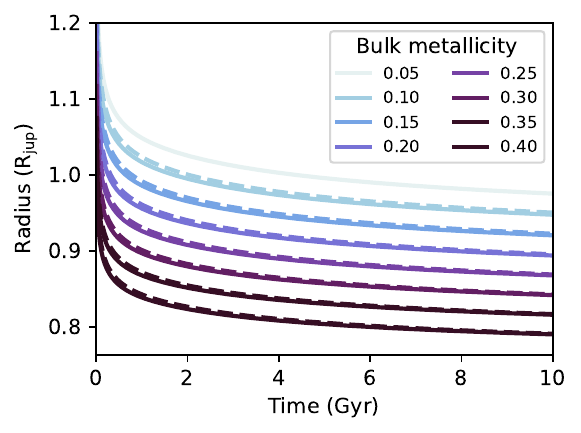}
    \caption{Planetary radius as a function of time for different bulk metallicities (coloured lines; see the legend). The solid lines are calculated from evolution models. The dashed lines use the heuristic that each 1\% change in radius can accommodate about a 2\% change in bulk metallicity (see the main text for details).}
    \label{fig:appendix_heuristic}
\end{figure}

\section{Radius evolution for different envelope metallicities}\label{sec:appendix_evolution}

{Figures \ref{fig:appendix_evolution_2}, \ref{fig:appendix_evolution_4}, and \ref{fig:appendix_evolution_10} show additional cooling tracks of planets with super-solar envelope metallicities of 2, 4, and 10 times solar. As discussed in Sect. \ref{sec:z_env}, for $T_{\rm{eq}} = 100$ K higher metallicities suppress deep radiative zones, which can be seen in the small differences between the models with different $w$ values. This is no longer true for the models with $T_{\rm{eq}} \gtrsim 400$ K, where once again there are significant differences in the cooling tracks for different $w$. These results further illustrate that also giant exoplanets with super-solar atmospheric metallicities could harbour deep radiative zones. We note that the $M = 0.3$ M$_{\rm{J}}$ model with $T_{\rm{eq}} = 100$ K and $w = 0.00, 0.90$ was extrapolated for ages beyond about 4 Gyr due to numerical issues.}

\begin{figure}[ht]
    \centering
    \includegraphics[width=0.95\linewidth]{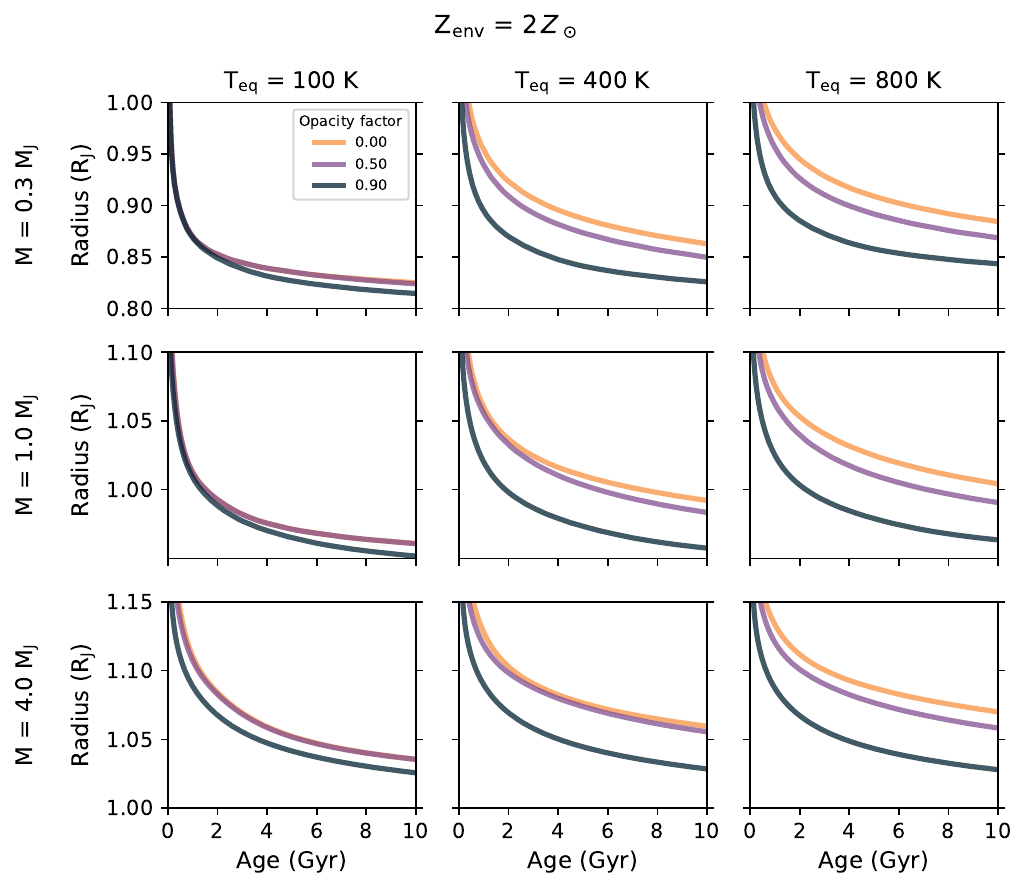}
    \caption{Radius evolution for different planetary masses (\textit{rows}), equilibrium temperatures (\textit{columns}), opacity factors (coloured lines; see the legend), and envelope metallicity of two times solar.}
    \label{fig:appendix_evolution_2}
\end{figure}

\begin{figure}
    \centering
    \includegraphics[width=0.95\linewidth]{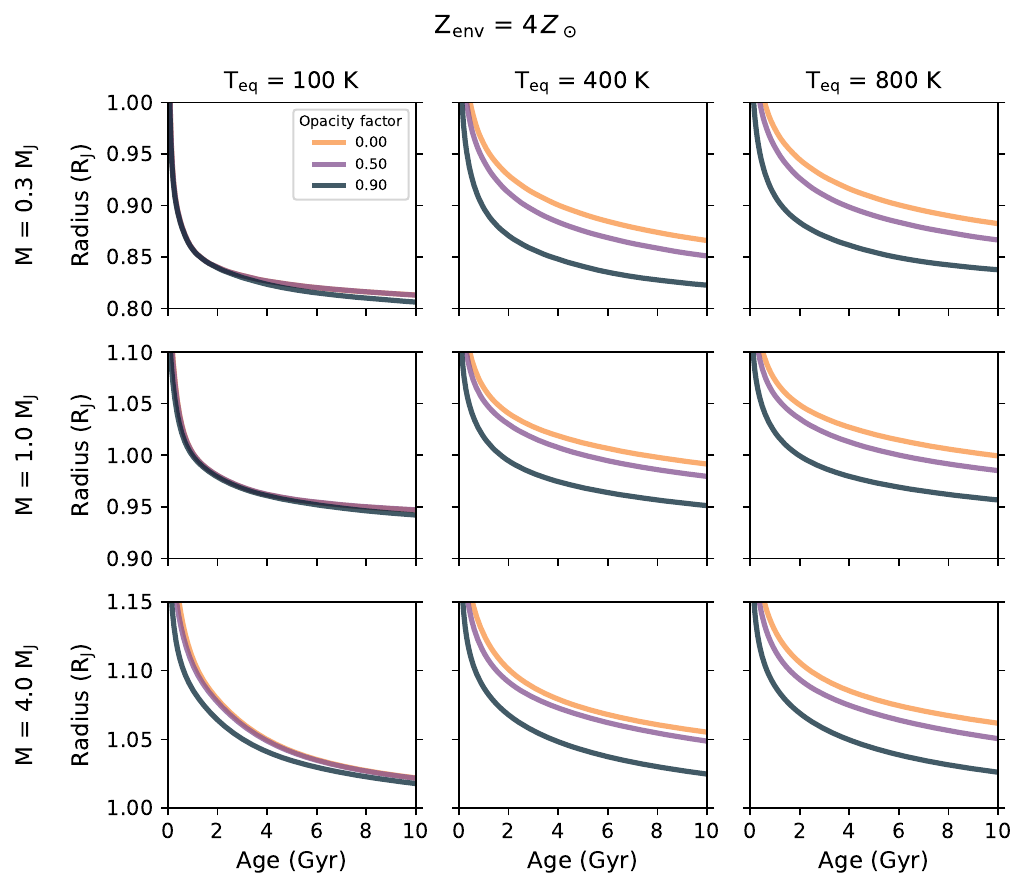}
    \caption{Same as Fig. \ref{fig:appendix_evolution_2} but for an envelope metallicity of four times solar.}
    \label{fig:appendix_evolution_4}
\end{figure}

\begin{figure}
    \centering
    \includegraphics[width=0.95\linewidth]{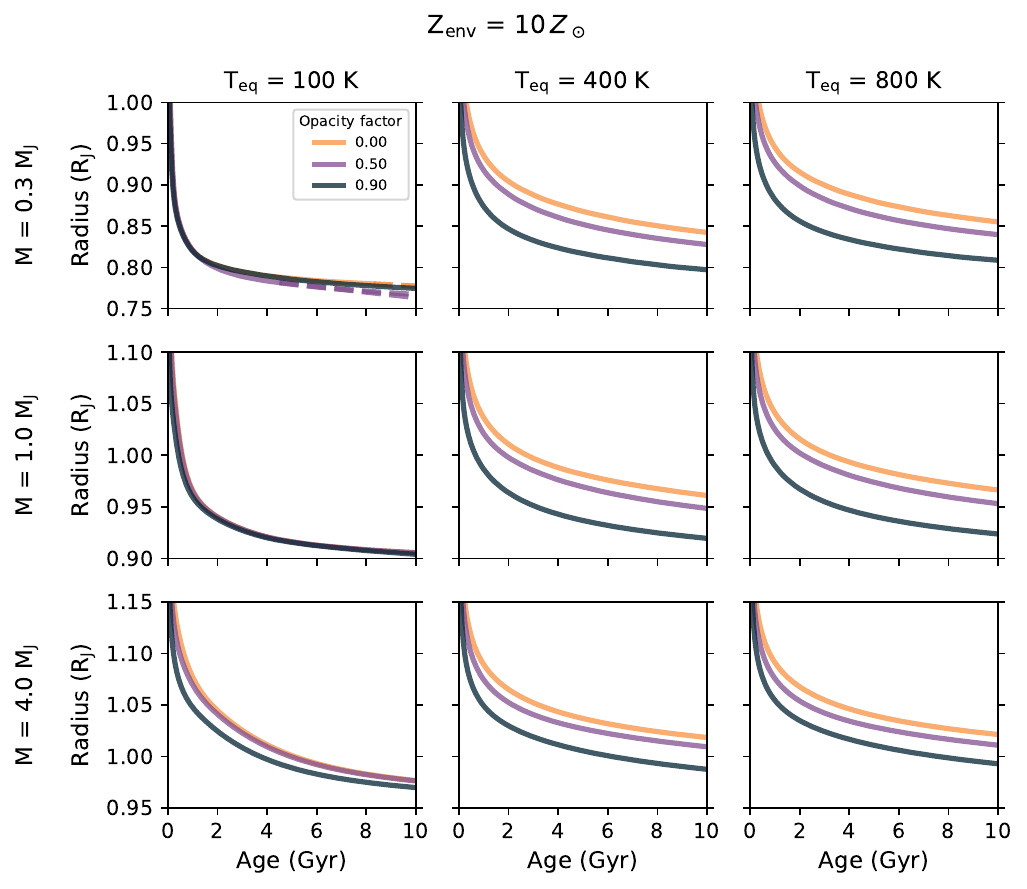}
    \caption{Same as Fig. \ref{fig:appendix_evolution_2} but for an envelope metallicity of ten times solar. The dashed evolution curves were extrapolated since the models did not converge after about 4 Gyr.}
    \label{fig:appendix_evolution_10}
\end{figure}

\onecolumn
\section{Additional models comparing the influence of the equilibrium temperature}\label{sec:appendix_teq}
{Figure \ref{fig:appendix_kippenhahn_teq} shows similar results to the ones  presented in Sect. \ref{sec:t_eq}, but for three additional masses of 0.6, 1.5, and 2.0 M$_{\rm{J}}$. The model with 1.5 M$_{\rm{J}}$ clearly shows the existence of a critical equilibrium temperature for which a deep radiative zone appears even when the opacity is unmodified. In this particular case, it is around $T_{\rm{eq}} = 150$ K, but this value strongly depends on the assumed mass and envelope metallicity.}

\begin{figure}[ht]
    \begin{subfigure}[b]{\linewidth}
        \centering
        \includegraphics[width=0.65\linewidth]{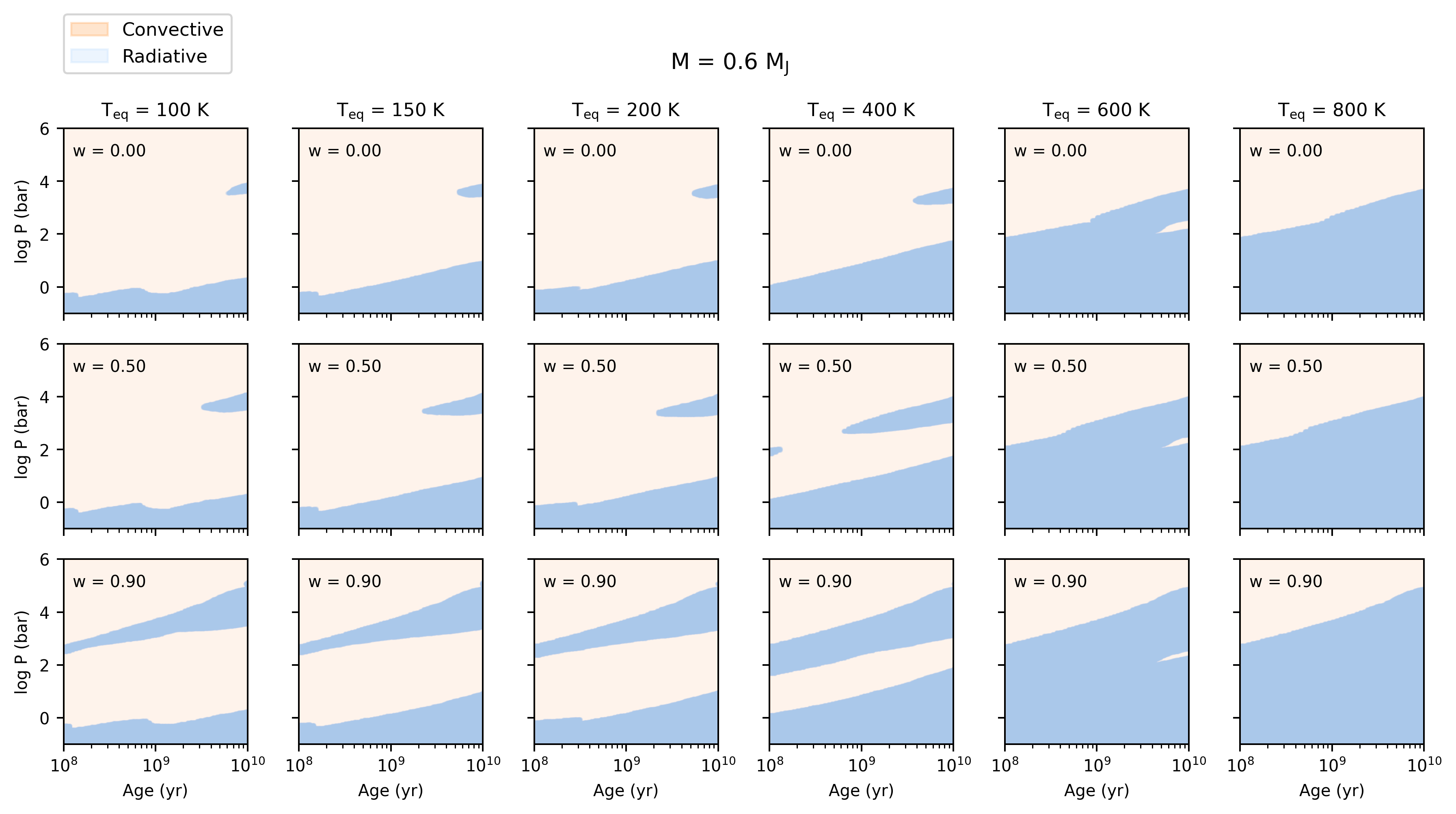}
    \end{subfigure}
    \begin{subfigure}[b]{\linewidth}
        \centering
        \includegraphics[width=0.65\linewidth]{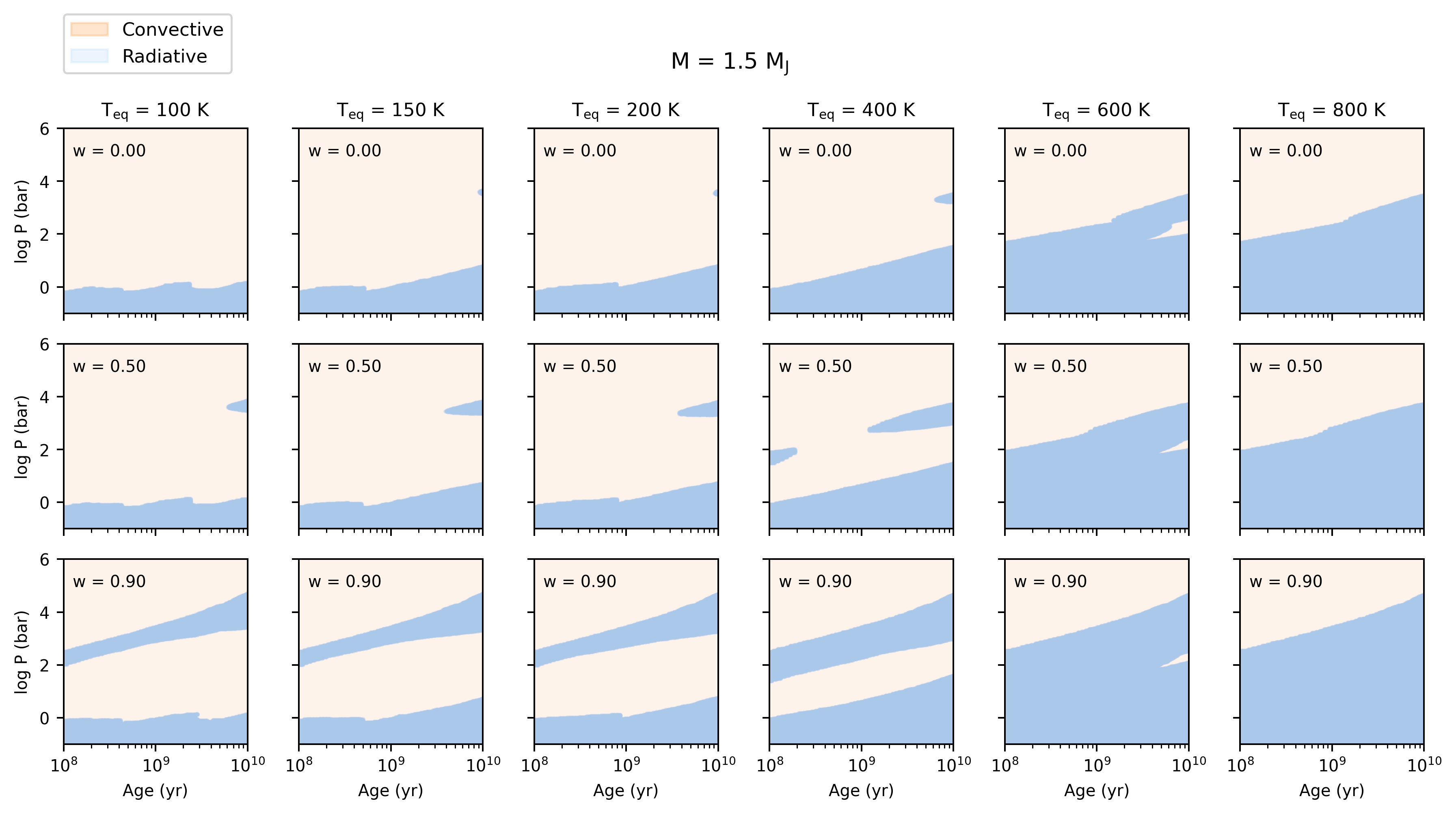}
    \end{subfigure}
    \begin{subfigure}[b]{\linewidth}
        \centering
        \includegraphics[width=0.65\linewidth]{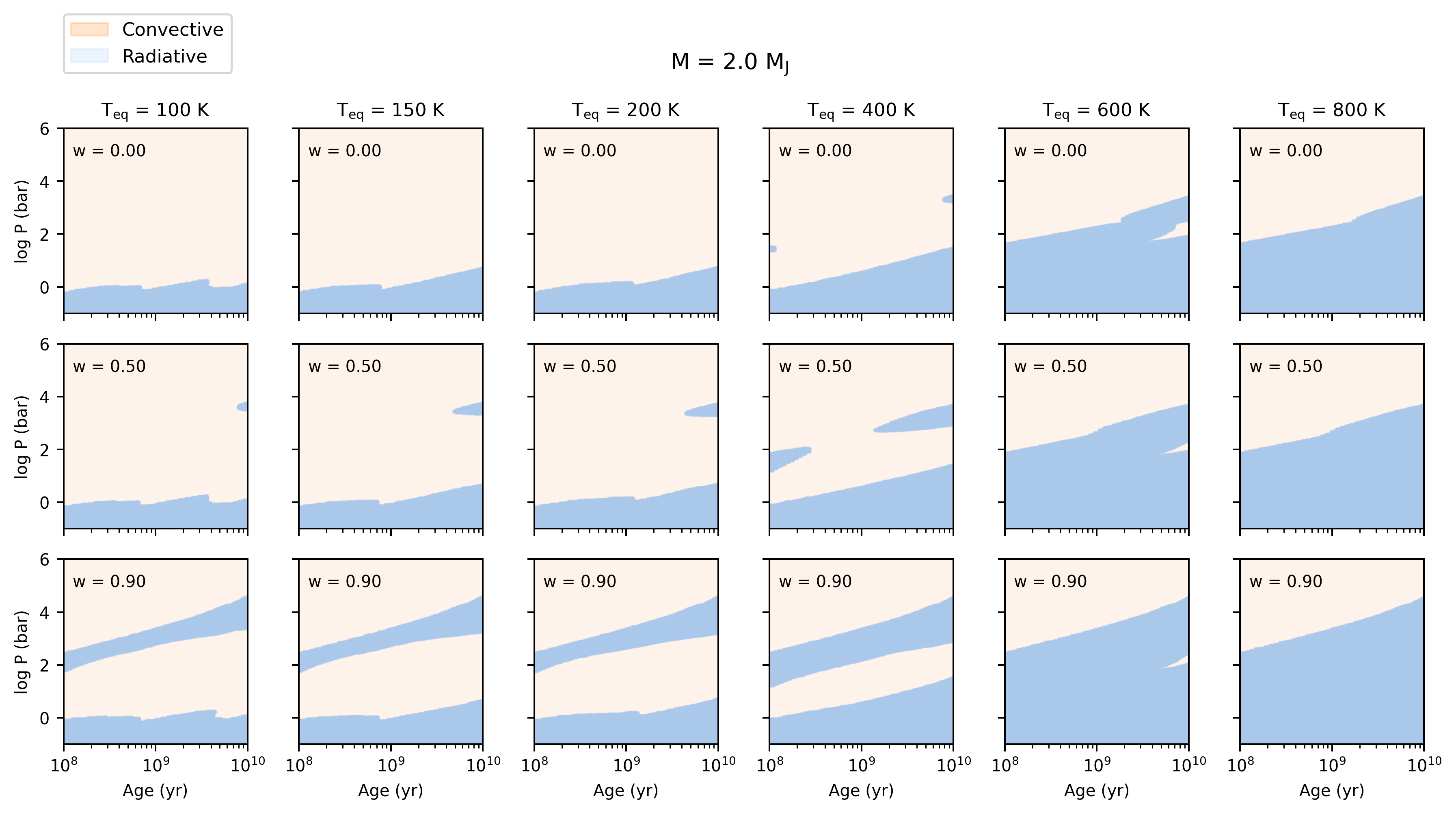}
    \end{subfigure}
    \caption{{Kippenhahn diagrams showing the evolution of convective (orange) and radiative zones (blue) with time for planetary masses of $M = 0.6$ (\textit{top}), $1.5$ (\textit{centre}), and $2.0$ M$_{\rm{J}}$ (\textit{bottom}). The columns and rows correspond to different equilibrium temperatures and opacity factors.}}
    \label{fig:appendix_kippenhahn_teq}
\end{figure}

\onecolumn
\section{Additional models comparing the influence of the envelope metallicity}\label{sec:appendix_zenv}

\begin{figure*}[ht]
    \centering
    \begin{subfigure}{0.49\linewidth}
        \centering
        \includegraphics[width=\linewidth]{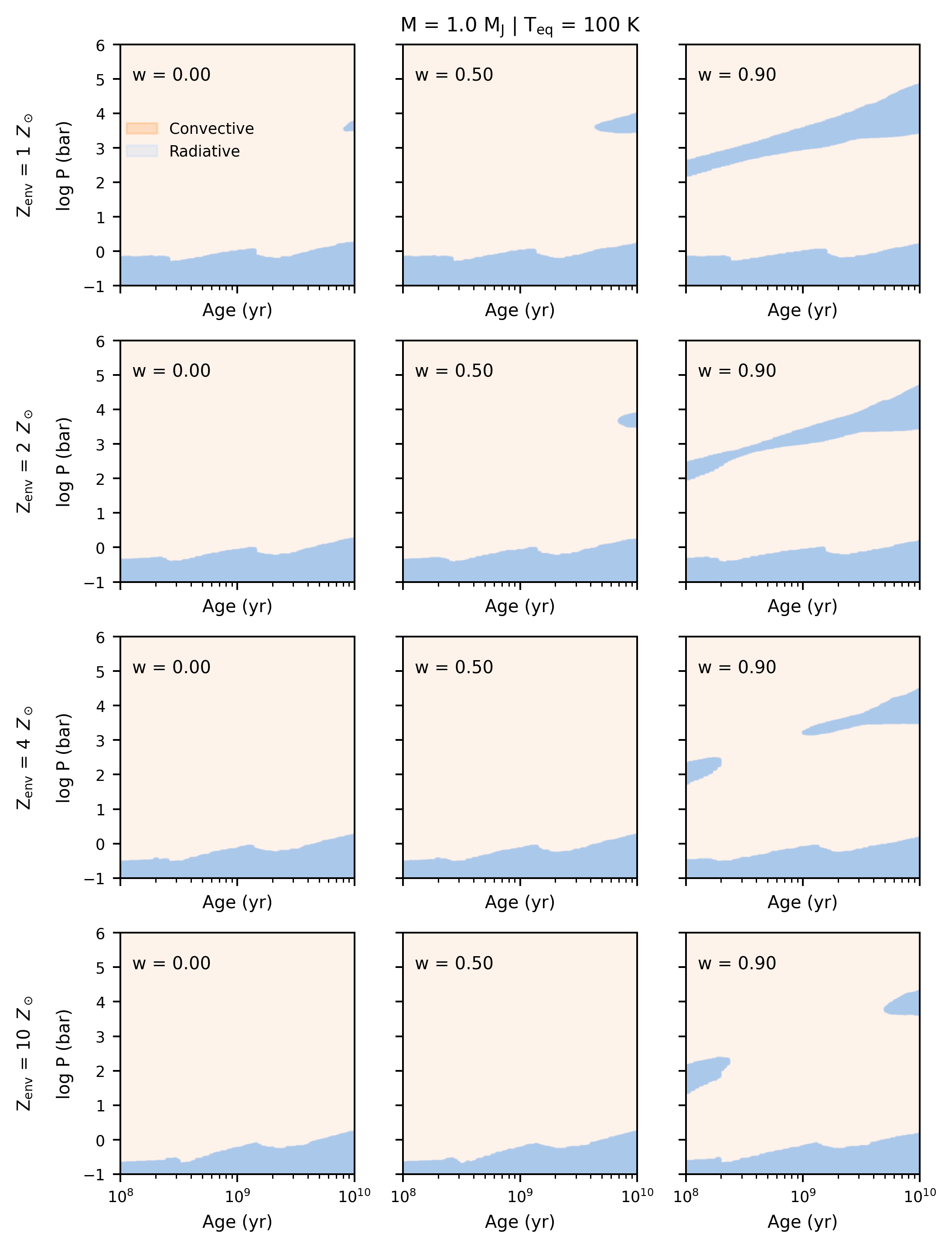}
    \end{subfigure}
    % \hfill
    \begin{subfigure}{0.49\linewidth}
        \centering
        \includegraphics[width=\linewidth]{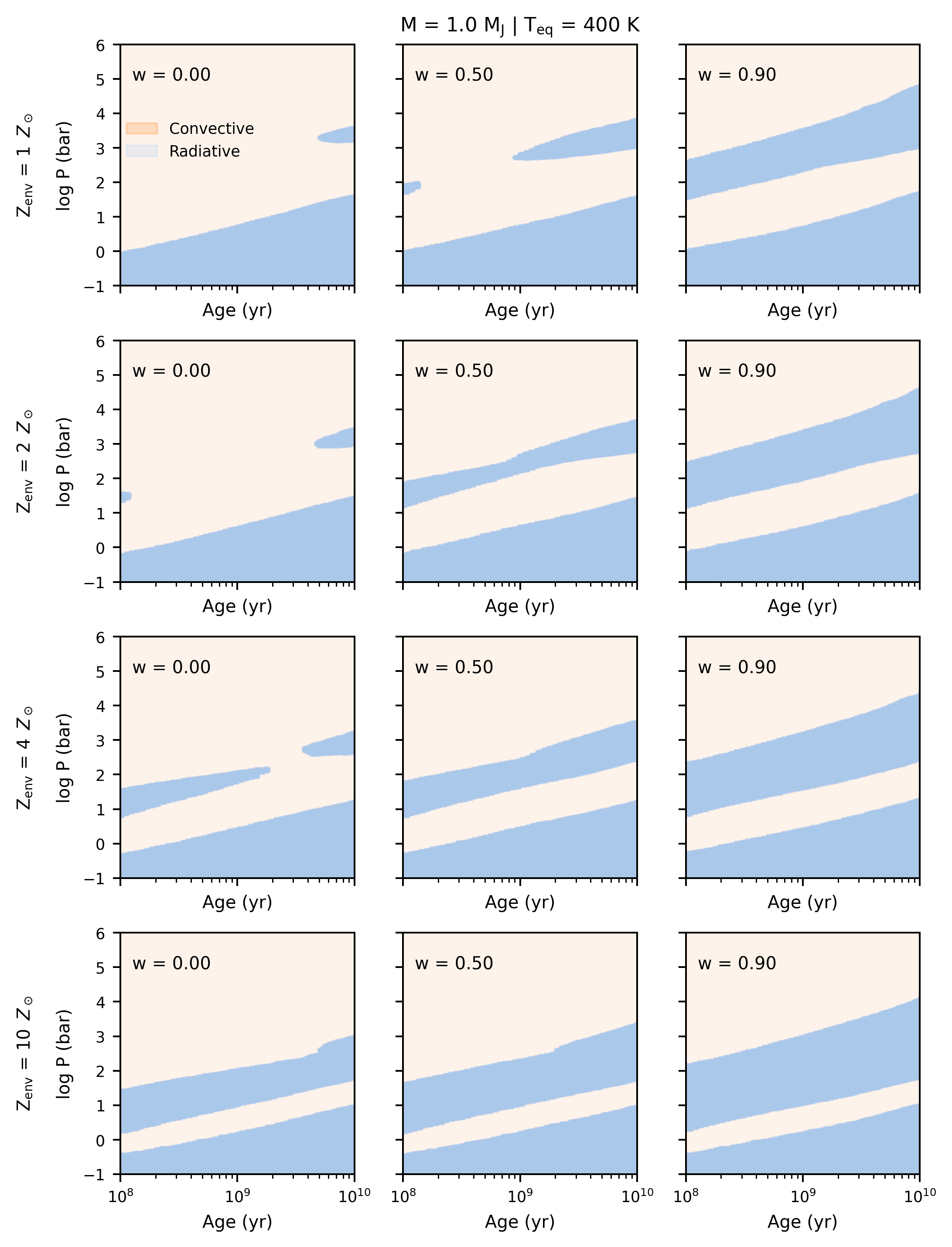}
    \end{subfigure}
    \caption{{Same as Fig. \ref{fig:kippenhahn_zenv_03} but for a planetary mass of $M = 1.0$ M$_{\rm{J}}$.}}
    \label{fig:kippenhahn_zenv_10}
\end{figure*}

\begin{figure*}[ht]
    \centering
    \begin{subfigure}{0.49\linewidth}
        \centering
        \includegraphics[width=\linewidth]{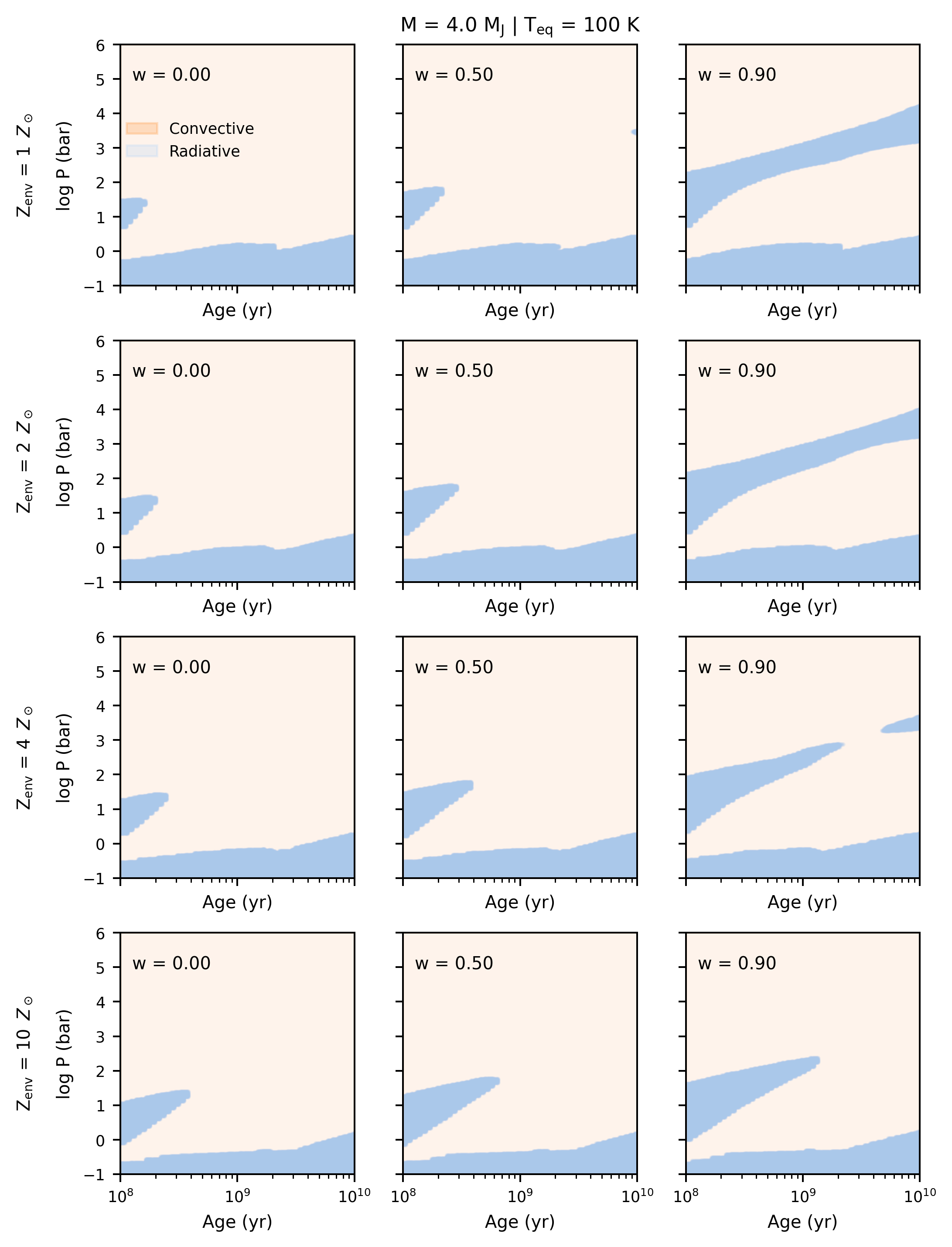}
    \end{subfigure}
    \begin{subfigure}{0.49\linewidth}
        \centering
        \includegraphics[width=\linewidth]{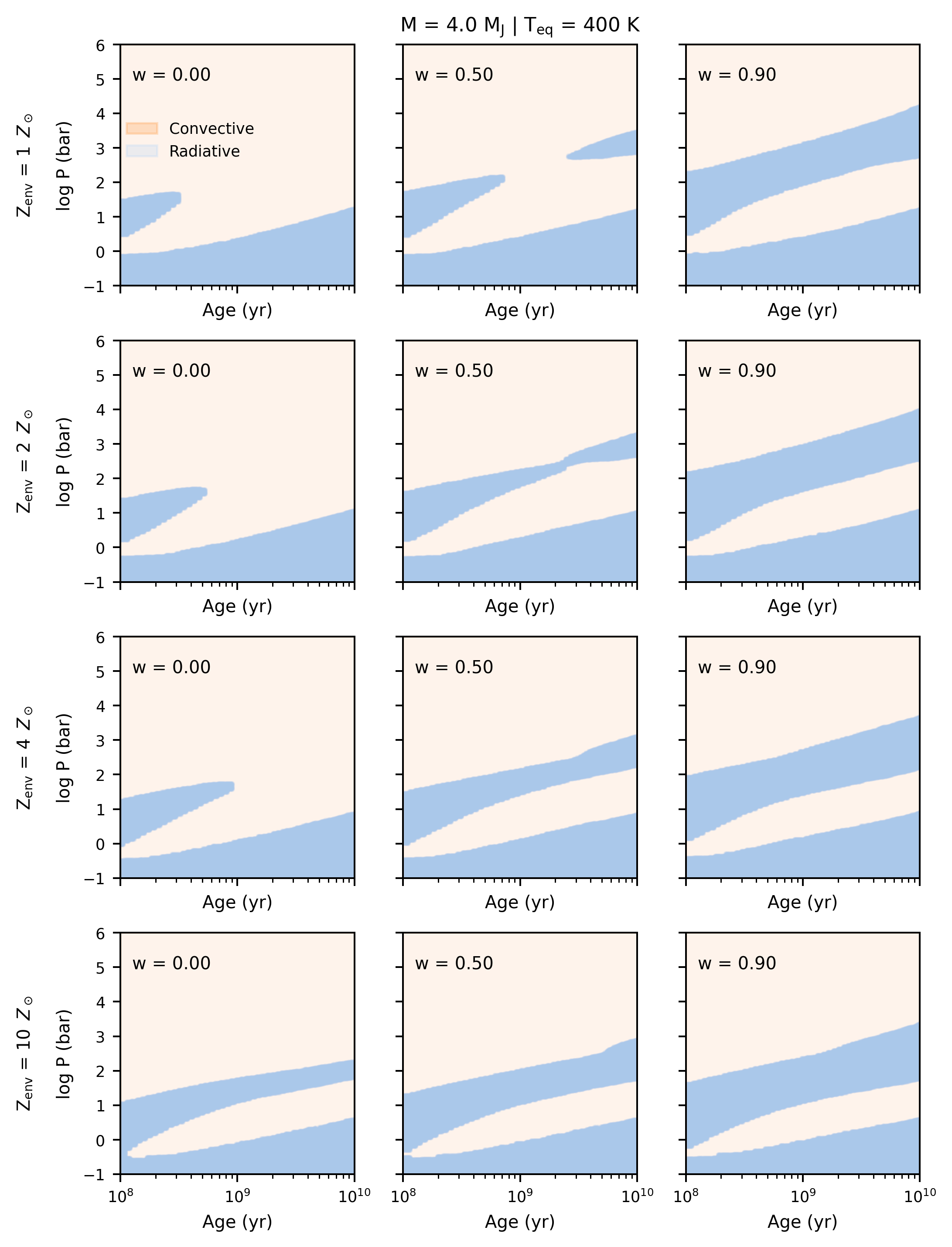}
    \end{subfigure}
    \caption{{Same as Figs. \ref{fig:kippenhahn_zenv_03} and \ref{fig:kippenhahn_zenv_10} but for a planetary mass of $M = 4.0$ M$_{\rm{J}}$.}}
    \label{fig:kippenhahn_zenv_40}
\end{figure*}

\end{appendix}

\end{document}